\pdfoutput=1

\documentclass[11pt,twoside,a4paper,note,final]{cms-tdr}
\def\svnVersion{1885a0c-D}\def\svnDate{2019/07/16}

\begin{document}

\hyphenation{had-ron-i-za-tion}
\hyphenation{cal-or-i-me-ter}
\hyphenation{de-vices}

\newskip{\cmsinstskip} \cmsinstskip=0pt plus 4pt
\newskip{\cmsauthskip} \cmsauthskip=16pt

\thispagestyle{empty} 

\vspace*{-25mm}

{\large\sffamily
\hspace*{\fill}
\parbox{45mm}%
{
{\sffamily         {CERN-EP-2020-054}}                  
{\sffamily         {2020/04/07}}       
\\[2mm]
}
}

\vspace*{\fill}

\begin{center}
\LARGE
{\sffamily\bfseries Further studies on the physics potential} \\[3mm]
{\sffamily\bfseries  of an experiment using LHC neutrinos} \\[3mm]
\end{center}

\vspace*{\fill}
\small

N.~Beni$^{1}$$^{,}$$^{2}$, 
M.~Brucoli$^{2}$, 
V.~Cafaro$^{3}$,
F.~Cerutti$^{2}$,
G.M.~Dallavalle$^{3}$,
S.~Danzeca$^{2}$,  
A.~De~Roeck$^{2}$,
A.~De~R\'ujula$^{4}$,
D.~Fasanella$^{2}$,
V.~Giordano$^{3}$,
C.~Guandalini$^{3}$, 
A.~Ioannisyan$^{2}$$^{,}$$^{5}$,
D.~Lazic$^{6}$,
A.~Margotti$^{3}$,
S.~Lo~Meo$^{3}$$^{,}$$^{7}$, 
F.L.~Navarria$^{3}$,
L.~Patrizii$^{3}$,
T.~Rovelli$^{3}$, 
M.~Sabat\'e-Gilarte$^{2}$$^{,}$$^{8}$,
F.~Sanchez~Galan$^{2}$,
P.~Santos~Diaz$^{2}$,
G.~Sirri$^{3}$,
Z.~Szillasi$^{1}$$^{,}$$^{2}$,
C.-E.~Wulz $^{9}$
\\ [3mm]
\footnotesize
$^{1}$ Hungarian Academy of Sciences, Inst. for Nuclear Research (ATOMKI), Debrecen, Hungary\\
$^{2}$ CERN, CH-1211 Geneva 23, Switzerland \\
$^{3}$ INFN sezione di Bologna and Dipartimento di Fisica dell' Universit\`a, Bologna, Italy \\
$^{4}$ Inst. de Estructura de la Materia, Consejo Superior de Investigaciones Cientificas (CSIC), Madrid\\
$^{5}$ A.Alikhanyan National Science Laboratory, Yerevan Physics Institute, Yerevan, Armenia\\
$^{6}$ Boston University, Department of Physics, Boston, MA 02215, USA \\
$^{7}$ ENEA Research Centre E. Clementel, Bologna, Italy \\
$^{8}$ Universidad de Sevilla, Spain\\
$^{9}$ Institute of High Energy Physics of the Austrian Academy of Sciences, and Vienna University of Technology, Vienna, Austria\\

\normalsize

\vspace*{\fill}

\begin{center}
{\sffamily\bfseries Abstract }  \\[2mm]

\parbox[c]{1.0\textwidth}{

We discuss an experiment to investigate neutrino physics 
at the LHC in Run~3, 
with emphasis  on tau flavour.
As described in our previous paper~\cite{XSEN1},
the detector can be installed in the decommissioned TI18 tunnel, $\approx$480 m downstream the ATLAS cavern, after the first bending dipoles of the LHC arc. 
In that location, the prolongation of the beam Line-of-Sight from Interaction Point IP1 to TI18 traverses about 100 m of rock. 
The detector intercepts the intense neutrino flux, generated by the LHC beams colliding in IP1, at large pseudorapidity $\eta$, 
where neutrino energies can exceed a TeV. \\
This paper  focuses on optimizing global features of the experiment, like  detector mass and acceptance. 
Since the $\nu$N interaction cross section grows almost linearly with energy, the detector can be light and still collect a considerable sample of neutrino events; in the present study it weighs less than 3 tons. 
The detector is positioned off the beam axis,  slightly above the ideal prolongation of the LHC beam from the straight section, covering 
$7.4<\eta<9.2$.
In this configuration,  the flux at high energies (0.5-1.5 TeV and beyond) is found to be dominated by neutrinos originating directly 
from IP1, mostly from charm decays, 
of which $\approx$50\% are
electron neutrinos and $\approx$5\% are tau neutrinos.
The contribution of pion and kaon decays to the muon neutrino flux  is studied by means of 
simulations that embed  the LHC optics and found small at high energies.\\
The above studies indicate that 
with 150~fb$^{-1}$ of delivered LHC luminosity in Run~3
the experiment can record  
a few thousand very high energy neutrino charged current interactions
and over 50 tau neutrino charged current events.
}
\end{center}

\vspace*{\fill}

\parbox[t]{1.0\textwidth}{
\footnotesize
}

\parbox[t]{1.0\textwidth}{
\footnotesize
}
\normalsize

\clearpage

\pagestyle{plain}
\setcounter{page}{1}\pagenumbering{roman}

\tableofcontents

\clearpage

\setcounter{page}{1}\pagenumbering{arabic}

\section{Introduction}

Intense fluxes of neutrinos of all flavours
are produced
in collisions of the LHC proton beams at the IP1 and IP5 Interaction Points, respectively hosting the ATLAS and the CMS detectors.
At large pseudorapidities $\eta$, neutrinos attain TeV energies,
a new domain much beyond the energy of available neutrino data
from laboratory experiments
\cite{XSEN1, PDG}.
At very high energy, astrophysical measurements  exist
\cite{IceCube}, 
with limited amounts of data.

LHC offers a unique opportunity for  probing 
neutrino physics 
for E$_{\nu}$ larger than 300~GeV and up to a few TeV. 
Tau flavour neutrinos are especially interesting, since there are hints of deviations from the Standard Model in the third generation,
in the measurement of the ${\it W}$ decay branching ratio to $\tau$ at LEP
\cite{LEP}
and in measurements of the semileptonic decays of ${\it B}$ to ${\it D}$ and ${\it D^{*}}$
\cite{HFLAV}.

 \begin{figure}[b]
  \centering
    \includegraphics[width=0.55\textwidth]{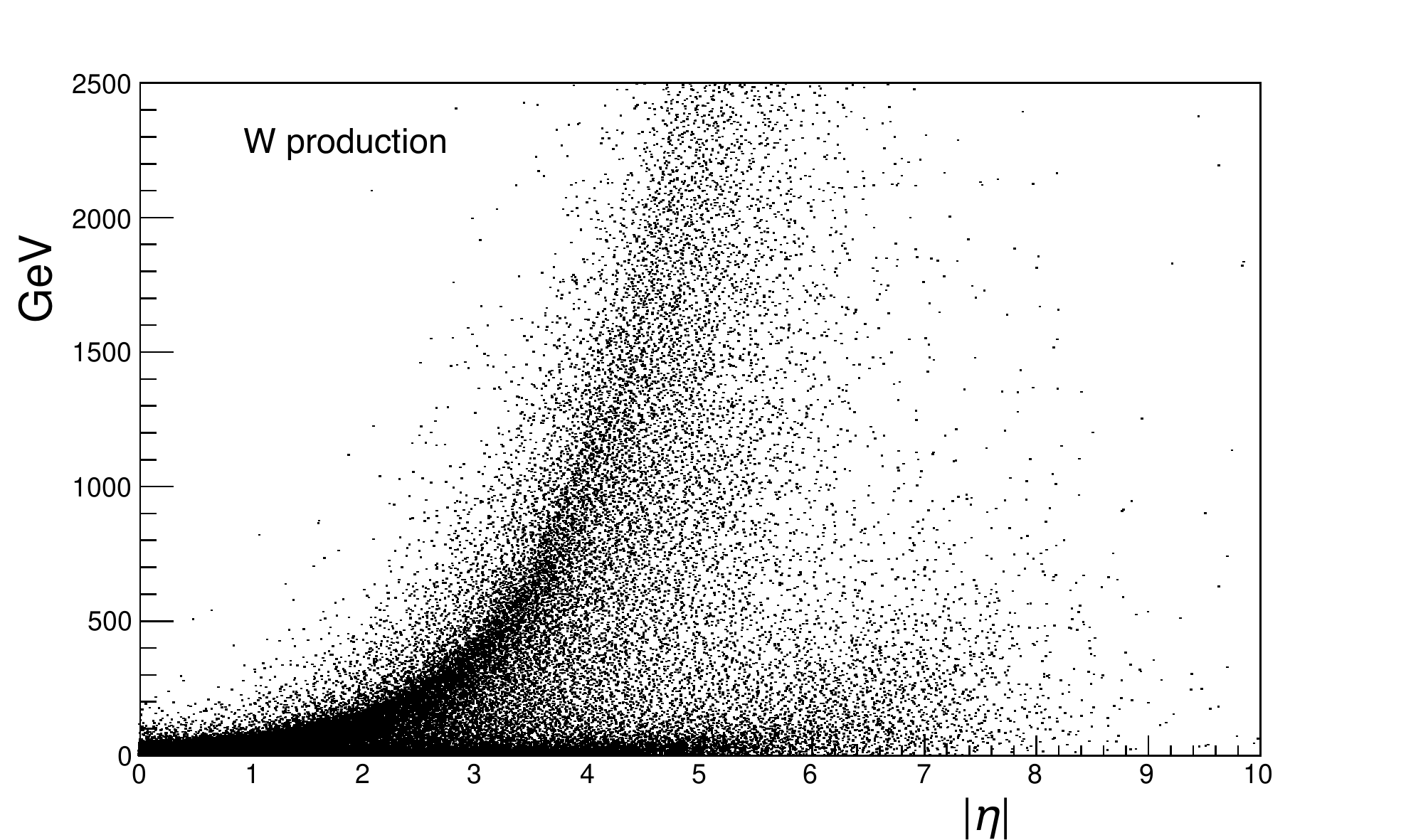}
   \includegraphics[width=0.55\textwidth] {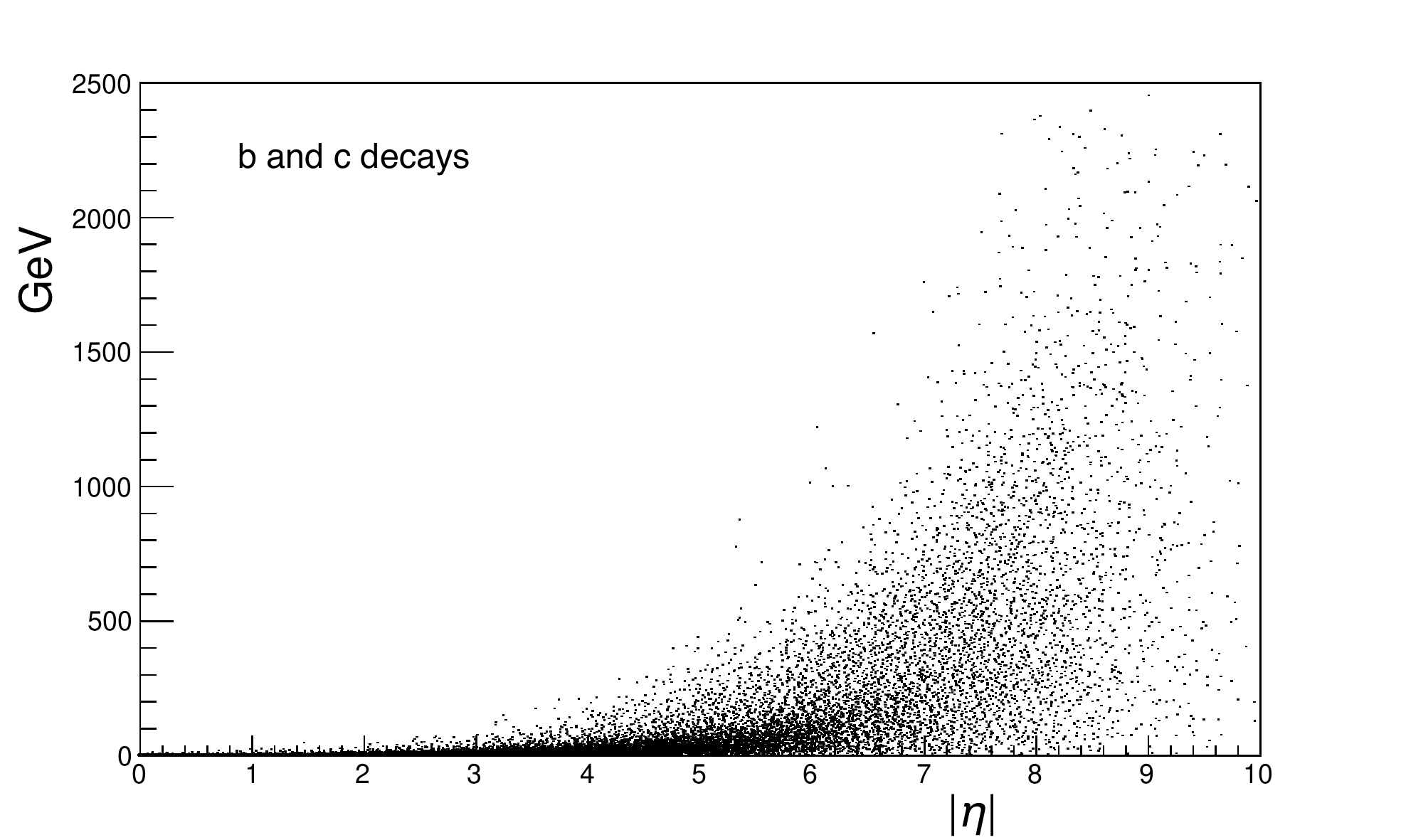}
\caption[Caption scatter plot Wbc] {Scatter plots of neutrino energy versus pseudorapidity $\eta$ 
in pp collision events
\cite{XSEN1}. 
Events generated 
with PYTHIA \cite{Pythia}.
Top: neutrinos from leptonic and hadronic W decays. 
Bottom: neutrinos from b and c decays. 
\label{fig:scatterWbc}}
\end{figure}

In a previous paper  we investigated  the feasibility of a neutrino experiment at the LHC
\cite{XSEN1}. 
We focused on high energy neutrinos in two $\eta$ ranges
(Figure~\ref{fig:scatterWbc}): 
\begin{itemize}
\item
$4<\eta<5$: high energy neutrinos are produced  in leptonic W decays;
about 33\% tau neutrinos
\item
$6.5<\eta<9.5$: neutrinos from c and b decays; about 5\% tau neutrinos, 
mostly from D$_{s}$ decays.
\end{itemize}

Four potential sites were identified and studied on the basis of
(a) expectations  for neutrino interaction rates, flavour composition and energy spectrum,  (b) predicted backgrounds and in-situ measurements, performed with a nuclear emulsion detector and radiation monitors.
One of the locations appeared to qualify for hosting a neutrino experiment. 
It lies at 480~m from the ATLAS IP, where the
cavern of the TI18 tunnel  intercepts the prolongation of the beam axis, after the beginning of the LHC arc.  That tunnel, now in disuse, was employed  to inject electron beams in LEP. A small mass detector located there can observe neutrinos from the IP within polar angles $<2.5$ mrad. A similar tunnel (TI12) exists at the opposite end of ATLAS. No such tunnels are present near the CMS IP.

In this paper we investigate global features and physics reach of an experiment located in TI18, with acceptance in $7.4<\eta<9.2$ and a mass of $<3$ tons, to take data during  the LHC Run~3. 
Simulations are performed using PYTHIA version~8
\cite{Pythia}
and the LHC FLUKA package of CERN EN-STI
\cite{Fluka1, Fluka2, DPMJET, LHCsim}, with embedded LHC optics.
For simplicity we assume that the detector is totally passive, made of emulsions and lead absorbers,
as used for our tests. However, the presented arguments apply to any detectors with similar mass and acceptance. 
Focus is on characteristics of the neutrino flux within acceptance, separating neutrinos produced 
during the pp collisions and those originating later 
in pion and kaon decays. 
Event rate expectations are calculated by neutrino flavour and
take into account only the detector mass; 
matters related to neutrino identification are detector specific and are not discussed.

\section{Detector Location}
 
A first requirement for the experiment feasibility is  a location in which particles coming directly from the IP are screened off, except for  muons and neutrinos,  by rock, or by the absorbers that protect  experimental areas and LHC components.  
A second requirement is that the local backgrounds from secondary interactions in collimators, beam pipe and other machine elements are low.
Intensity and composition of these machine induced backgrounds vary rapidly along the LHC, however they are well predicted by simulations performed with FLUKA, as extensive in-situ measurements have proven
\cite{XSEN1, Cerutti1, Cerutti2}.

\begin{figure} [b]
\centering
\includegraphics[width=0.7\textwidth]{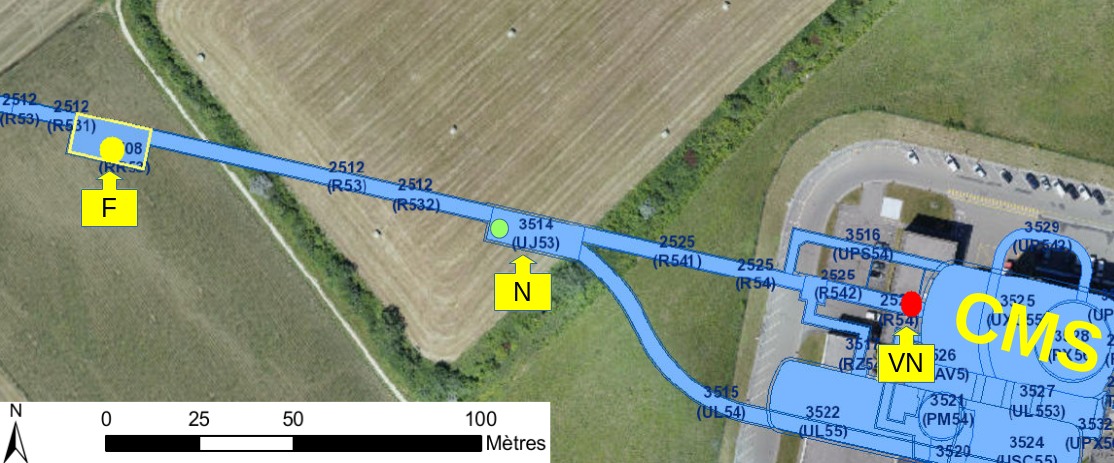}
\caption[Caption emulsions] {View of the VN (Q1 magnet), N (UJ53 hall), and F (RR53 hall) test locations showing their positions and distances from IP5, along the CMS LHC straight section
\cite{XSEN1}. 
\label{fig:CMS_VN_N_F}}
\end{figure}

\begin{figure} [b]
\centering
\includegraphics[width=0.8\textwidth]{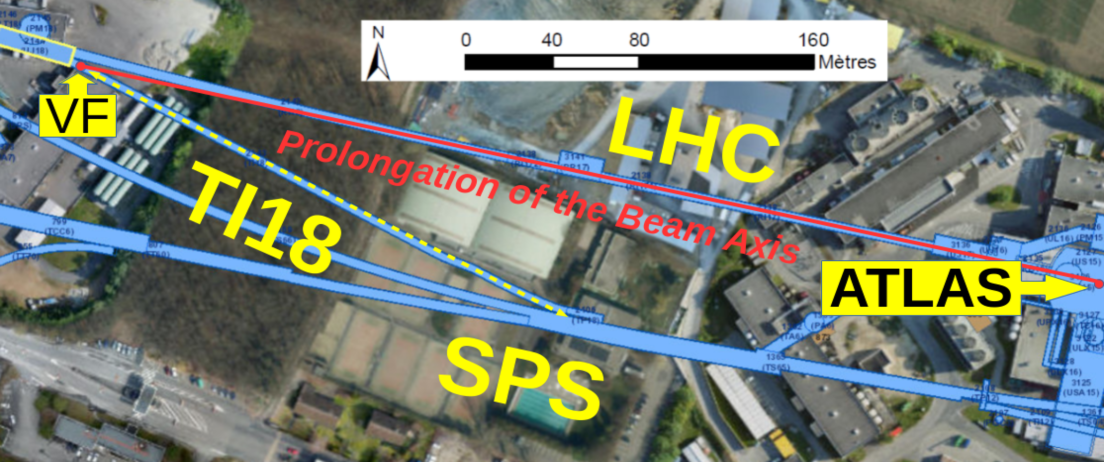}
\caption[Caption emulsions] {View of the VF location showing its position and distance from IP1. The Prolongation of the Beam Axis (PBA) of the ATLAS LHC straight section intercepts the cavern of the TI18 tunnel after the beginning of the LHC arc
\cite{XSEN1}. PBA is also called Line of Sight (LoS). \label{fig:ATLAS_VF_withPBA}}
\end{figure}

In reference 
\cite{XSEN1}
we studied four locations (named VN, N, F, VF in Figures
~\ref{fig:CMS_VN_N_F} and
~\ref{fig:ATLAS_VF_withPBA})
as potential hosts for a neutrino experiment:
the CMS inner triplet region (25~m from CMS IP), UJ53 and UJ57 (90 and 120~m from CMS IP), RR53 and RR57 (240~m from CMS IP), TI18 (480~m from ATLAS IP).

Nuclear emulsions, packaged in stacks with layers of lead as in the OPERA detector
\cite{OPERA}, were used in our tests because they do not require an infrastructure for detector services, and do not disturb the LHC magnets or other machine elements. 
The charged hadron fluence was independently measured also with CERN Radiation Monitors
\cite{radmon1, radmon2}, 
which complemented the emulsion package in the background tests.
The emulsions were protected from thermal neutrons by several cm thick layer of borated polyethylene.  

 In the three sites tested near IP5 (CMS) (Figure~\ref{fig:CMS_VN_N_F})
the muon fluence ranged from 1 to 6 $\times$10$^5$ /cm$^{2}$/fb$^{-1}$ in F and VN respectively,
the measured charged hadron fluence was about 10$^6$$-$10$^7$ /cm$^{2}$/fb$^{-1}$, and the thermal neutron fluence was 10$^7$$-$10$^8$ /cm$^{2}$/fb$^{-1}$ in the best location (F)
\cite{XSEN1}.
The measurements
were on average in agreement with the predictions 
 of the LHC machine simulations
\cite{Cerutti2}.

\begin{figure}[t]
  \centering
    \includegraphics[width=0.9\textwidth] {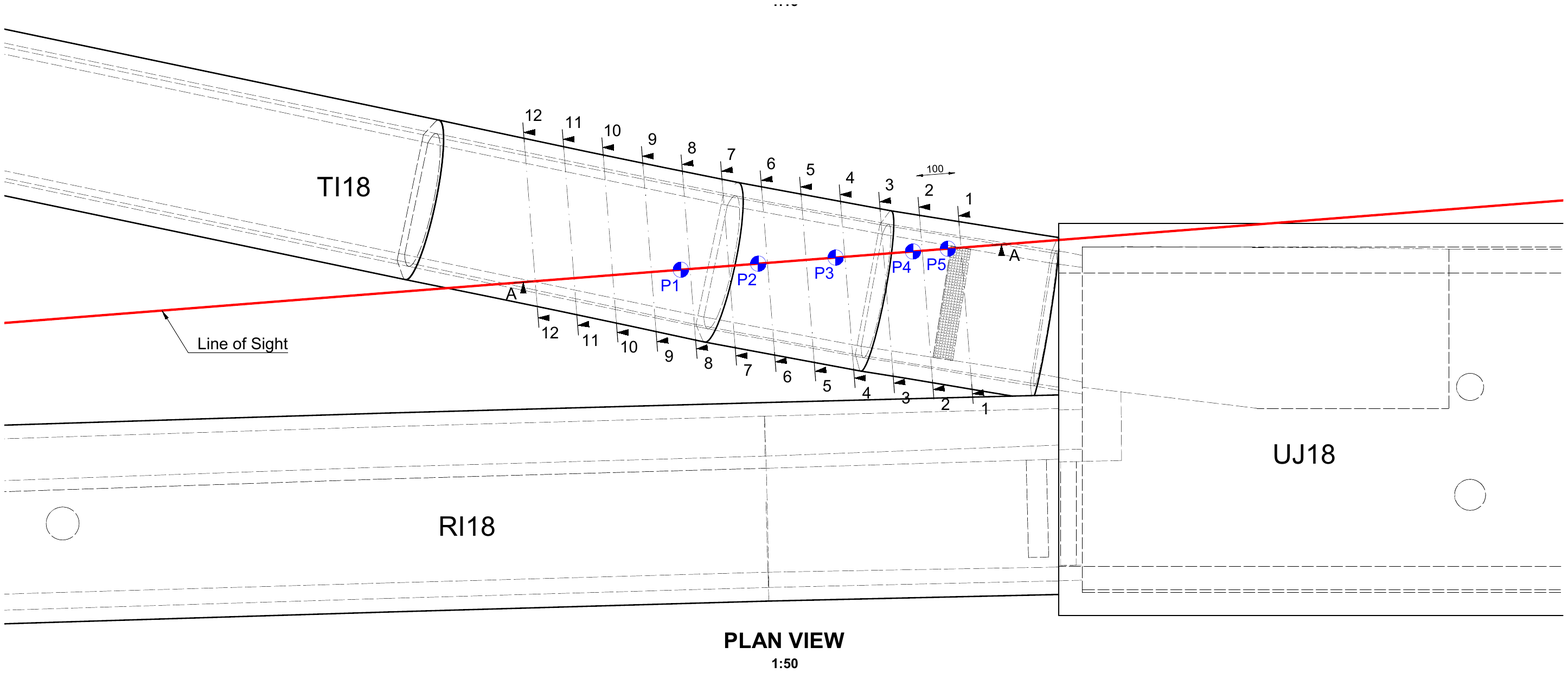}
    \includegraphics[width=0.7\textwidth]{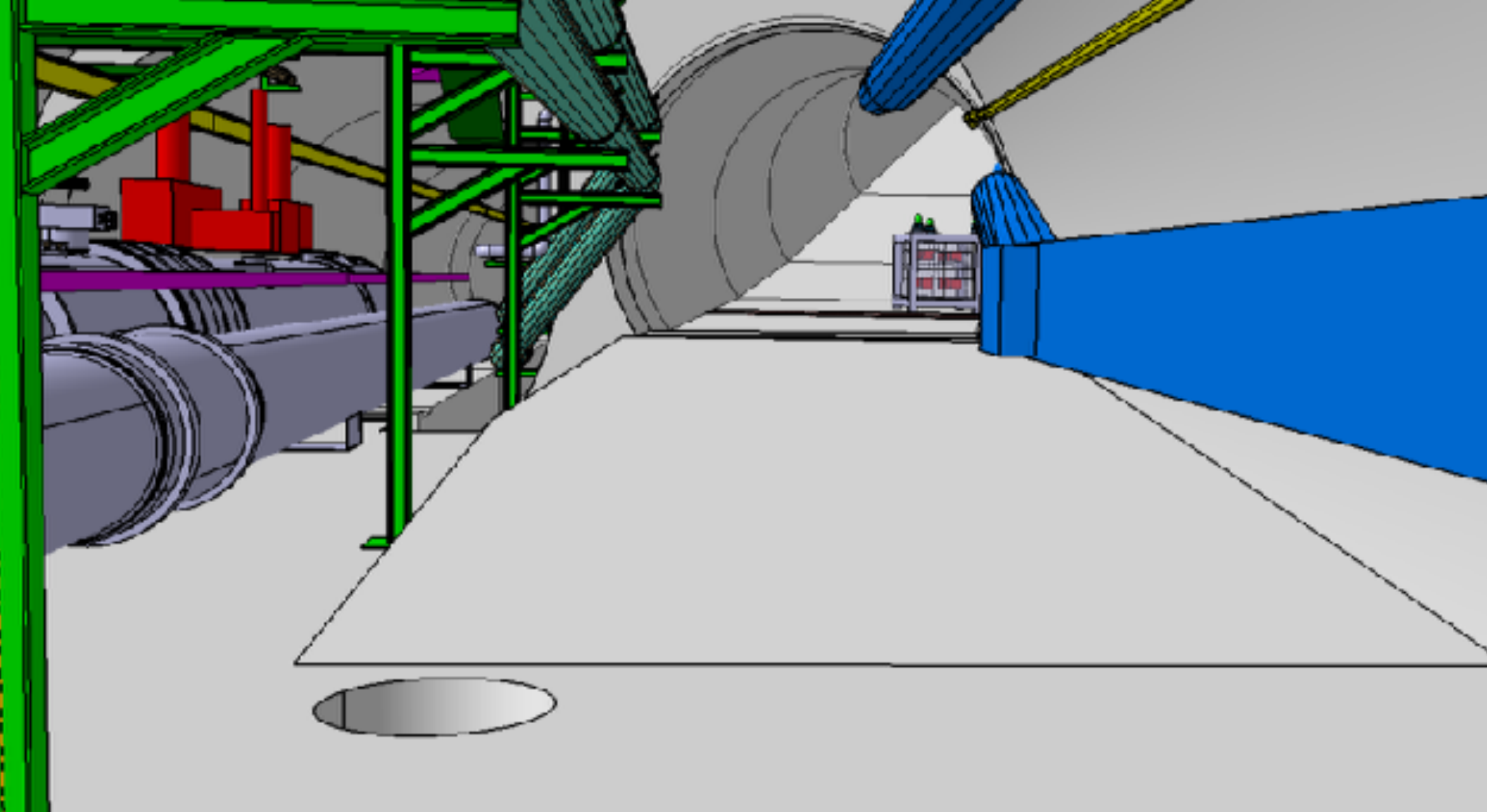}
\caption[Caption TI18] {TI18 tunnel at the connection with the LHC tunnel
\cite{CERN-EN}. Top: plan view, with survey marks (floor heights (P) along the LoS and tunnel contours at different heights); IP1 is 480~m on the left. In red the beam Line of Sight from IP1.
Bottom:  3d view.  
 \label{fig:TI18}}
\end{figure}

The VF site only exists near  the IP1 (ATLAS) region
(Figure~\ref{fig:ATLAS_VF_withPBA}). 
It exploits the decommissioned LEP injection tunnel TI18.
Survey measurements were performed by the CERN engineering and survey groups,
and were made available to us
\cite{CERN-EN}.
The cavern of the TI18 tunnel 
(Figure~\ref{fig:TI18})
intercepts the prolongation of the beam axis, named Line of Sight (LoS), at about 480 meters from the ATLAS IP, at the beginning of the collider's arc, downstream of the first bending dipoles.
The background levels in VF are at least an order of magnitude lower than in the best location (F) near CMS.

The muon flux from cosmic rays in TI18 is low. The cavern is 80.6 meters underground.  The flux can be evaluated from reference
\cite{cosmics}, considering a standard rock density of 2.65~g/cm$^3$;
 without restricting the direction of impact on the emulsions, we estimate an upper limit of  
30~tracks/cm$^2$/day. 
In average during Run~2, a few days were needed for the LHC to deliver 1~fb$^{-1}$ of luminosity.

In summary, the VF location in the TI18 tunnel qualified as a suitable host for a neutrino experiment.
Energetic charged particles are deviated by the LHC arc optics and do not reach the detector;
the beam LoS from IP1 traverses  about 100~meters of rock before crossing the TI18 cavern.
However, 
the useful space is restricted in length to a few meters and the floor is on a slope that lies higher than the LoS from a minimum of 5~cm growing in steps up to 50~cm:
these are limitations for a standard massive neutrino detector, but not in this case, as shown in the following.

\section{Experiment Generalities
\label{sec:ExpGen}}

\begin{figure} [t]
\centering
\includegraphics[width=0.7\textwidth]{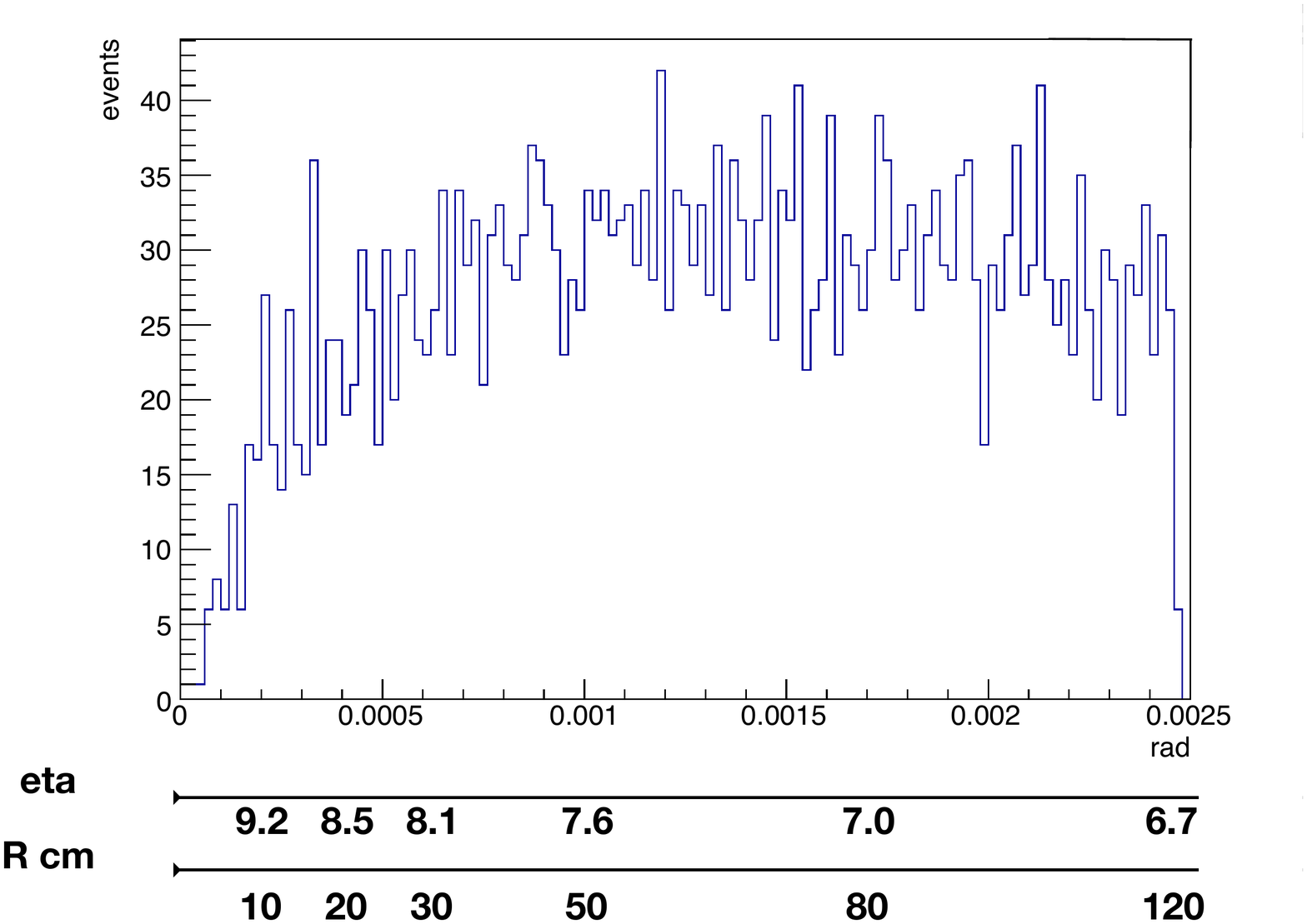}
\caption[neutrinotheta] {Polar angle distribution of neutrinos from b and c decays in the VF location ($\eta>$6.7). Events generated with PYTHIA
\cite{Pythia}.
Also indicated are the pseudorapidity $\eta$ range and the distance R perpendicular to the beam axis.
\label{fig:neutrinotheta}}
\end{figure}

In order to maximize the amount of tau neutrinos, the detector acceptance should favour the contribution from b and c decays.  
Figure~\ref{fig:neutrinotheta} 
shows the polar angle distribution of neutrinos from b and c decays; about 5\% of those are tau neutrinos:
a configuration with the detector slightly off the beam axis, although very close to it, is favoured.
 
The $\nu$N cross section grows rapidly with the neutrino energy, linearly from 10~GeV to a few hundred GeV 
\cite{PDG}. 
Therefore, at high energy the detector can be light, featuring a mass of a few tons, and still collect a considerable sample of neutrino interactions 
~\cite{XSEN1};
the energy spectrum of the observed events will be hard, because the higher energy neutrinos have larger interaction cross sections.
If lead is used as target, the detector becomes very compact: 
1~ton of lead is a block of 1~meter length and 30x30~cm$^2$ cross section.

Most neutrino experiments designed for tagging tau neutrino interactions embedded nuclear emulsions in their detectors
\cite{OPERA, donut, DsTau}. 
Emulsions are efficient for reconstructing the vertex of tau decays.
The OPERA experiment
\cite{OPERA} used a modular design: emulsions interleaved with thin layers of lead, packed together into  a “brick”.
An OPERA brick is 128~mm wide, 104~mm high, and 78~mm thick, with 56~mm lead; 
it weighs ~8.3~kg.
A 1~ton detector requires 120 such bricks.

Thus for the sake of the present studies, we will assume our detector consisting of lead-emulsion layers subdivided in OPERA-like bricks. 
Of course there are, in reality, important detector performance limitations, which are briefly recalled in the following, but they do not affect the arguments exposed in this paper, which apply to any detectors with similar mass and $\eta$ acceptance. 
  
In our background tests
\cite{XSEN1}, 
it was measured that emulsions could stand 10$^7$~tracks/cm$^2$.
However, the track extrapolation between emulsions across a lead layer becomes more complex when the track density is high;
a level of 10$^5$~tracks/cm$^2$ is regarded as a good condition.
Thus, given the background estimates in TI18,
it is essential  that emulsion exposure does not exceed 30~fb$^{-1}$. 
In view of the luminosity that LHC is expected to deliver during Run~3, 
it means replacement of the bricks a few times per year.

\begin{figure} [t]
\centering
\includegraphics[width=0.7\textwidth]{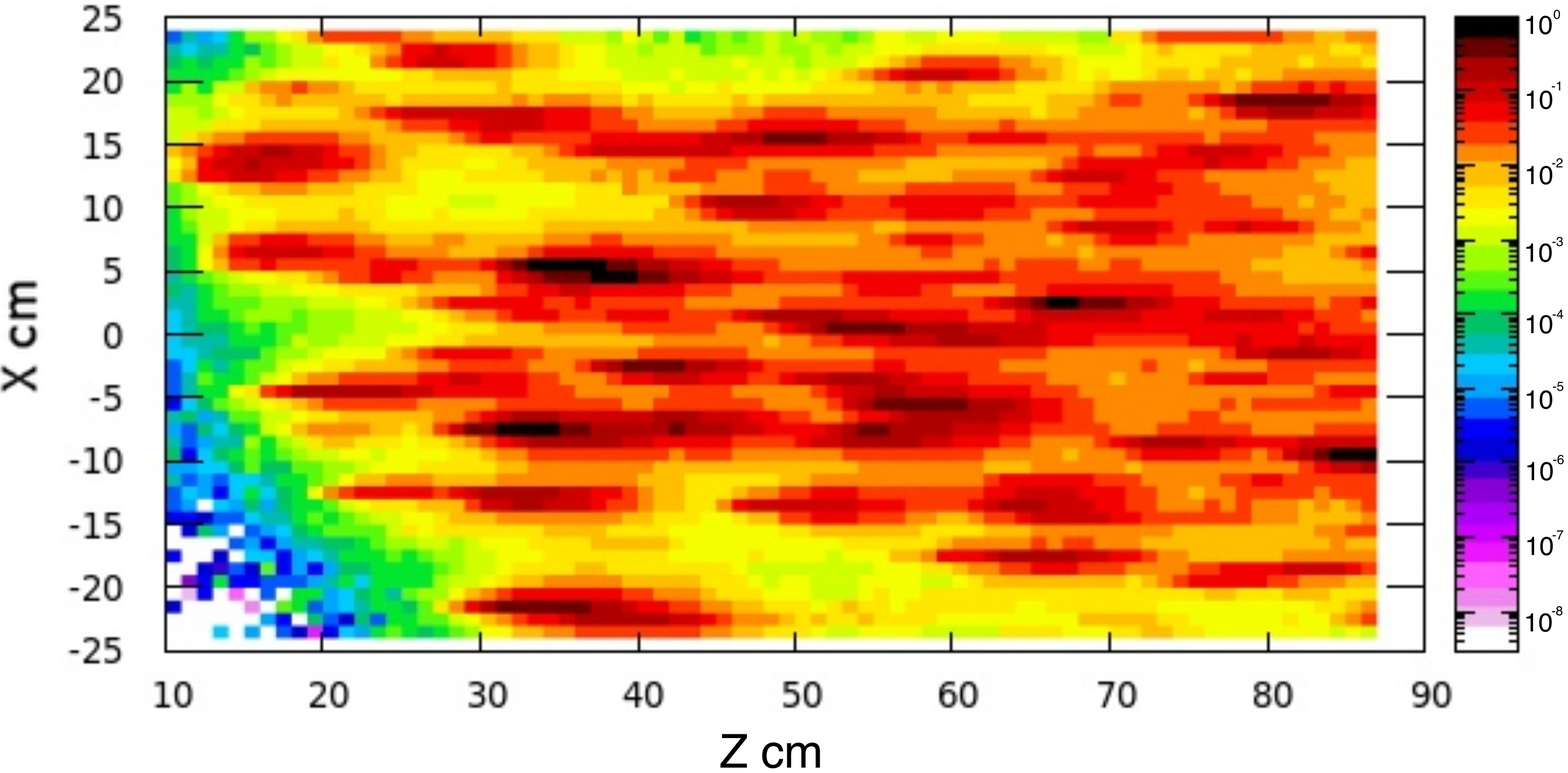}
\caption[Exz] { FLUKA
\cite{Fluka1, Fluka2}
simulation of hundred electron $\nu$, with $0.2<E<1.2$ TeV, interacting in a 77~cm deep structure of
OPERA-like lead-emulsion layers. 
Incoming neutrinos are randomly distributed over an area X,Y of  of 50x30 cm$^{2}$. 
Normalization is to the elementary volume with the largest energy deposit.  
\label{fig:Exz}}.
\end{figure}

Emulsion layers in an OPERA brick  are uniformly spaced: a one millimiter lead sheet separates two consecutive layers, for a total of 56~layers. 
The brick thickness  is short for reconstructing with high efficiency 
a decay vertex of a tau with TeV energy.
The expected decay length of a tau lepton is a few centimeters, much longer than in the OPERA case, suggesting that the structure should be extended to include at least 200 layers.
Besides, a TeV electron shower has a 95\% energy containement length of 22~cm, about 4 OPERA bricks, as observed in 
Figure~\ref{fig:Exz},
showing  hundred  charged current (CC) interactions of electron neutrinos with energies from 0.2 to 1.2 TeV, simulated with FLUKA
\cite{Fluka1, Fluka2}
in a structure like an OPERA brick but repeated ten times in depth (Z).

\begin{figure} [t]
\centering
\includegraphics[width=0.6\textwidth]{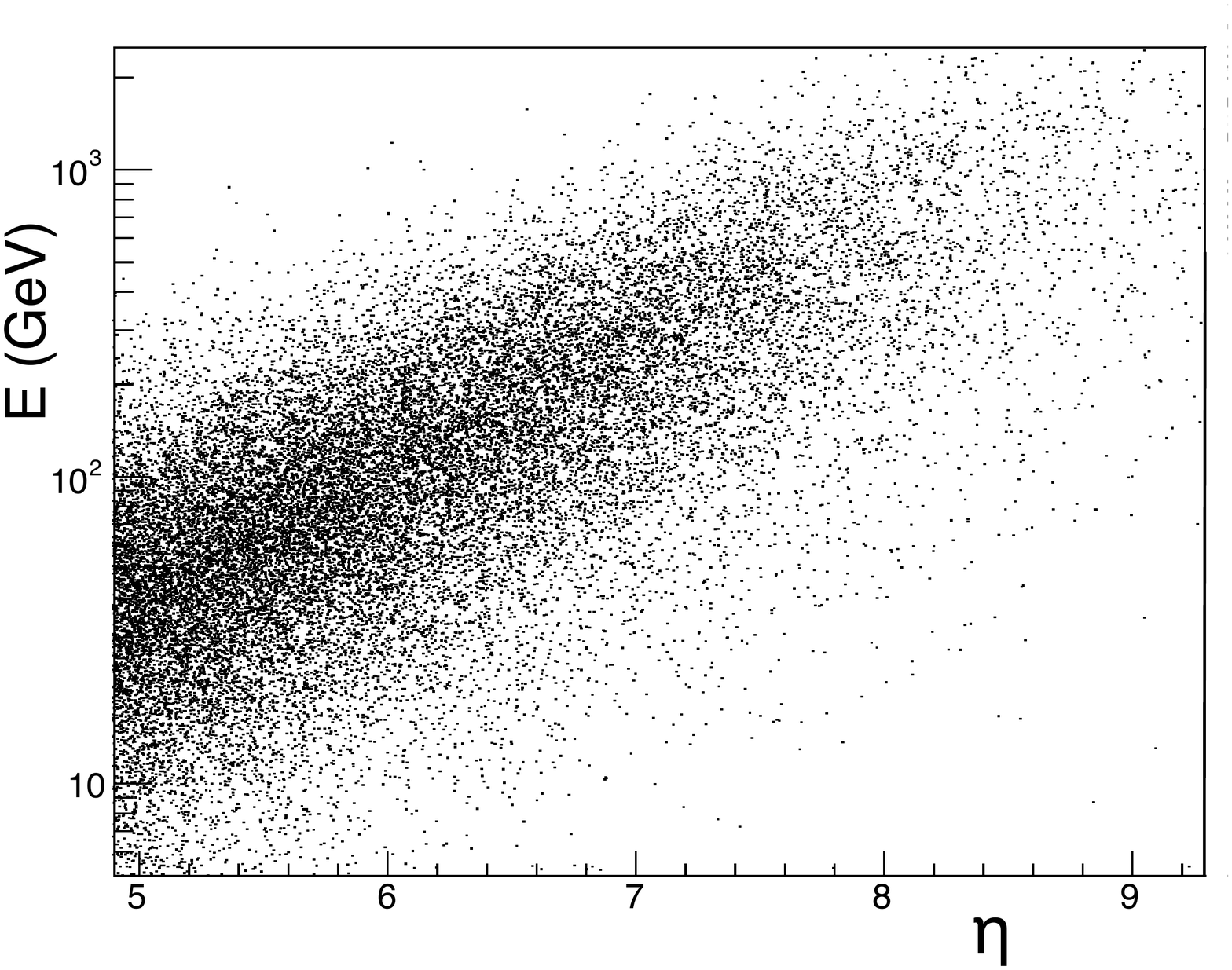}
\caption[neutrinotheta] {A scatter plot of log(E) versus $\eta$ for neutrinos from b and c decays.
Events generated with PYTHIA
\cite{Pythia}. 
\label{fig:logE_vs_largeeta}}.
\end{figure}

The emulsion stacks are very good for reconstructing the vertex of a neutrino interaction, but they cannot easily provide a measurement of the neutrino energy. 
On an event by event basis the neutrino energy can be estimated using the methods developed in OPERA and other emulsion based detectors, which for instance exploit the correlation between track multiplicity and neutrino energy
~\cite{OPERA_MCSrec, OPERA_EMrec, SHIP_EMrec}. 
The achievable resolution depends on the brick structure; a resolution of  50\%  in the 0.1-1 TeV energy range seems realistic.
However, also kinematics can be exploited: in the regime of longitudinal momentum p$_{L}$ much larger 
than transverse momentum p$_{T}$, the pseudorapidity of particles emerging from the IP is proportional to the logarithm of the energy, 
a relation smeared by the particle p$_{T}$ distribution.
Different $\eta$ ranges have different average energy, as seen in Figure~\ref{fig:logE_vs_largeeta}.
The  log($E_{\nu}$) versus $\eta$ scatter plot can also be useful for defining a fiducial phase space so to reject obvious outliers.

In the following we study a design with two detectors made of OPERA-like bricks, 
of $\approx$1~ton each, $\approx$1~m thick.
They cover different $\eta$ ranges, to be optimised for 
high energy and tau flavour neutrinos.
We do not aim for high resolution in energy, but rather for having two independent energy bins, one for each detector.

\begin{figure}[h]
  \centering
    \includegraphics[width=0.6\textwidth]{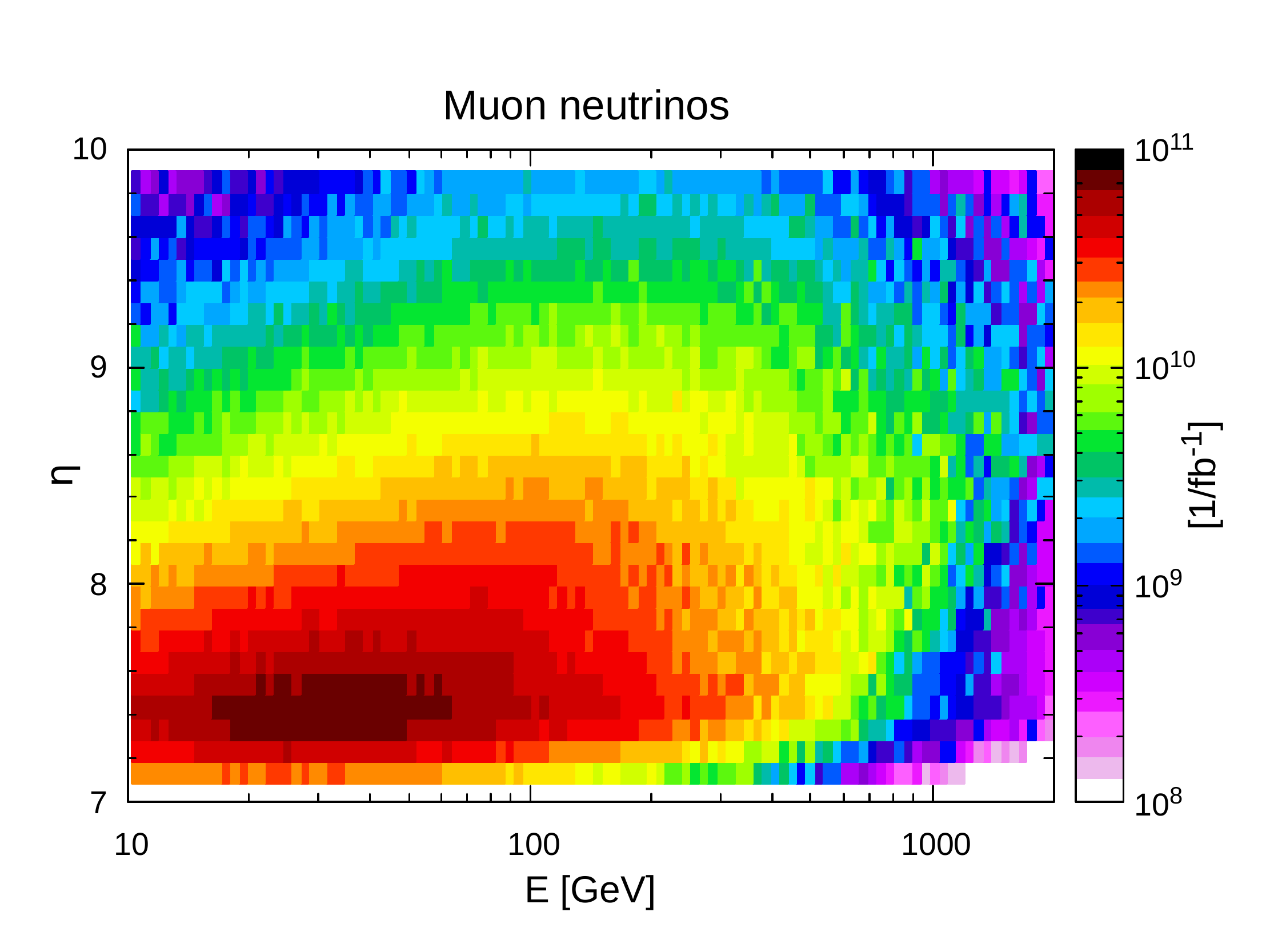}
   \includegraphics[width=0.6\textwidth] {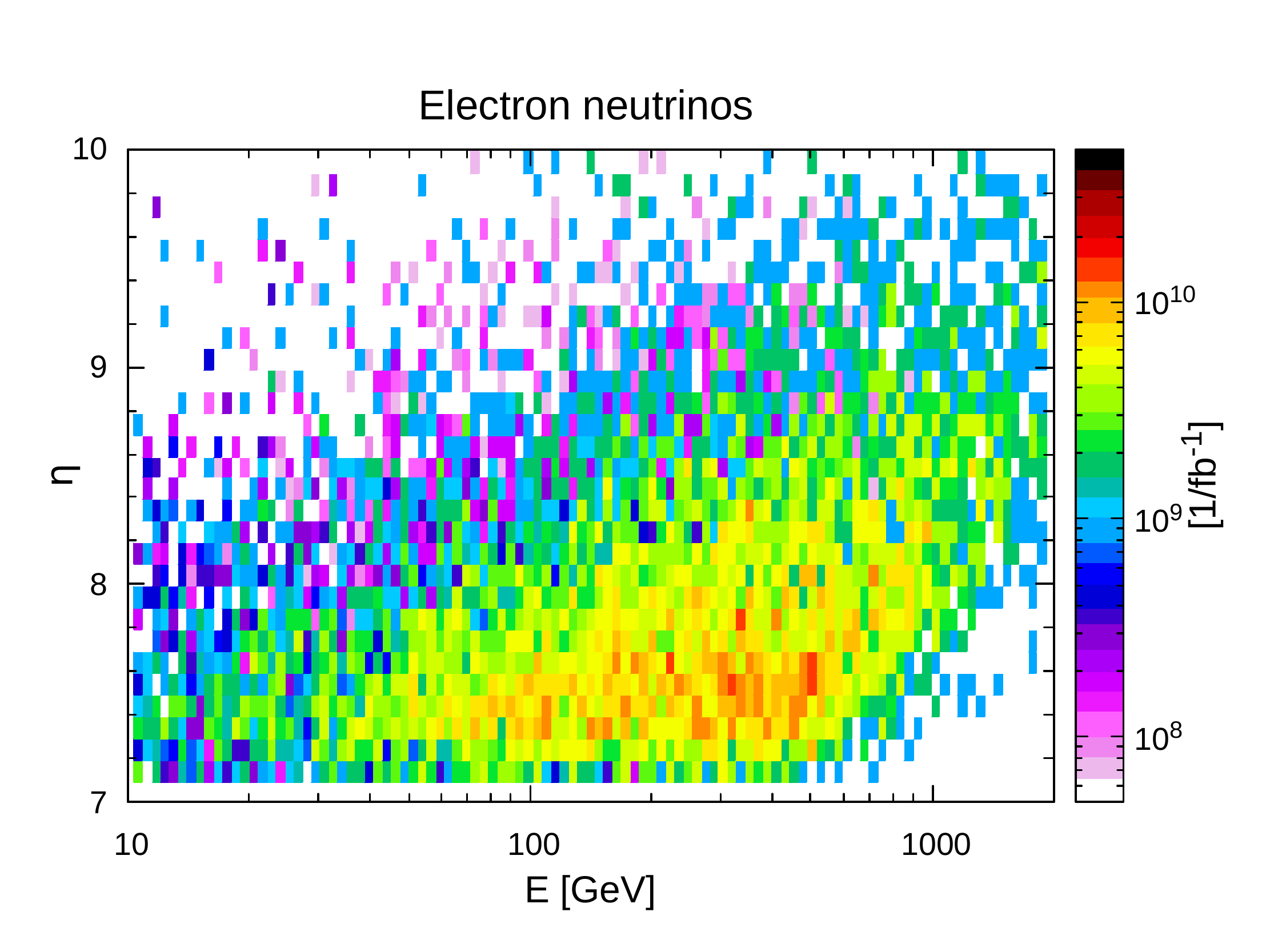}
\caption[Caption DPMJET_scatter] {Scatter plots of neutrino pseudorapidity $\eta$ versus energy. 
Events were generated with FLUKA 
using the embedded DPMJET event generator, and LHC simulation; both pion/kaon decays and charm production were included.
Top: muon neutrinos. Bottom: electron neutrinos. \label{fig:DPMJET_scatter}}
\end{figure}

\section{Neutrino Flux
\label{sec:NuFlux}}

PYTHIA 
\cite{Pythia}
was used 
to simulate proton-proton collision events at $\sqrt{s}=$ 13~TeV and to estimate the neutrino flux
in TI18.
At high energy  most of the flux  originates 
from  c ($\approx$92\%) and b ($\approx$8\%) decays; about 5\% of the neutrinos are of the tau flavour, originating in $D_{s}\rightarrow\tau\nu_{\tau}$ decays and in the subsequent $\tau$ decays.  
Neutrinos from pion decays pointing towards TI18 are predicted to have mostly low energies. 
The $\gamma$c$\tau$ of a pion exceeds 500~m already at 10~GeV, and  it reaches 5~km 
at 100~GeV. Therefore most high energy  pions are deviated by the LHC optics and interact in the LHC beam pipe or in the rock before they can decay.
Neutrinos from kaon decays pointing towards TI18 can have higher energies, 
since the $\gamma$c$\tau$  for a 100 GeV kaon is about 740 m, 
however the kaon/pion production rate ratio is only about 11\%.

PYTHIA was tested in depth by the LHC experiments; it reproduces the features of proton-proton interactions with good accuracy.
Measurements of charged particle production in the forward direction were performed by LHCb
\cite{LHCb_charged} in the pseudorapidity range $2.0<\eta<4.5$, and by TOTEM
\cite{TOTEM_charged}
 in  $5.3<\eta<6.5$, and found in reasonable agreement with PYTHIA expectations. 
However, no experimental crosschecks exist for the very forward $\eta$ range that our detector subtends.
We explored several possibilities for benchmarking our findings with PYTHIA.

Previous calculations of neutrino flux and CC event rates for a similar experiment 
were performed by
A. De R\`ujula, E. Fernandez and J.J. G\`omez-Cadenas 
\cite{derujula}~(1993)
and by H. Park
\cite{Park}~(2011).
De R\`ujula et al. made an analytical calculation using two variants of a non-perturbative QCD model (Quark Gluon String Model), which had large uncertainties in predicting the transverse momentum distribution for the produced hadrons.
Park used a PYTHIA generator version 6 with parameter set tuned to Tevatron data
and very early LHC data, and 
a lower \textit{D$_{s}$} production cross section.
The detector was assumed to be on the beam LoS, and the calculation shows a peak for low energy muon neutrinos very close to the beam axis ($\eta>9$).
Although these neutrinos are mostly outside our detector acceptance, their exact distribution at about 500~m from the IP critically depends on the transport of the parent pion/kaon along the LHC optics. 
We conclude that neither calculations can be used for benchmarking our PYTHIA predictions.

A study of the expected flux in TI18 very close the beam LoS 
(within 10~cm around it, i.e. $\eta>9.1$ )  was presented in a recent paper by the FASER collaboration
\cite{FASERnu},
but it is not directly applicable to
our detector, 
which covers a different $\eta$ range. 

Finally a good test was provided by the CERN FLUKA team in the EN-STI group 
that  carried out a thorough  study deploying an independent event generator with state-of-the-art LHC optics emulation. 
Proton-proton collisions were generated with FLUKA 
using the embedded DPMJET event generator, which  describes particle production in pp minimum-bias events, including charm production;
then pions and kaons, before decaying, were transported through LHC elements and environment material up to TI18
\cite{Fluka1, Fluka2, DPMJET, LHCsim}.
Information was stored separately for neutrinos and antineutrinos, and by flavour.
Figure~\ref{fig:DPMJET_scatter} shows the predicted fluence in $\eta$ versus energy, for the pseudorapidity range under consideration, for muon and electron neutrinos, in 1~fb$^{-1}$. 
Electron neutrinos show the $\eta$ vs ln(E) dependence of particles from the IP as in 
Figure~\ref{fig:logE_vs_largeeta}.
The distribution falls very rapidly at very large $\eta$ and it becomes inconvenient to move closer to the beam LoS.
In the following we split the study in two pseudorapidity regions: $7.4<\eta<8.1$, $8.0<\eta<9.2$.

\begin{figure}[htbp]
  \centering
    \includegraphics[width=0.55\textwidth]{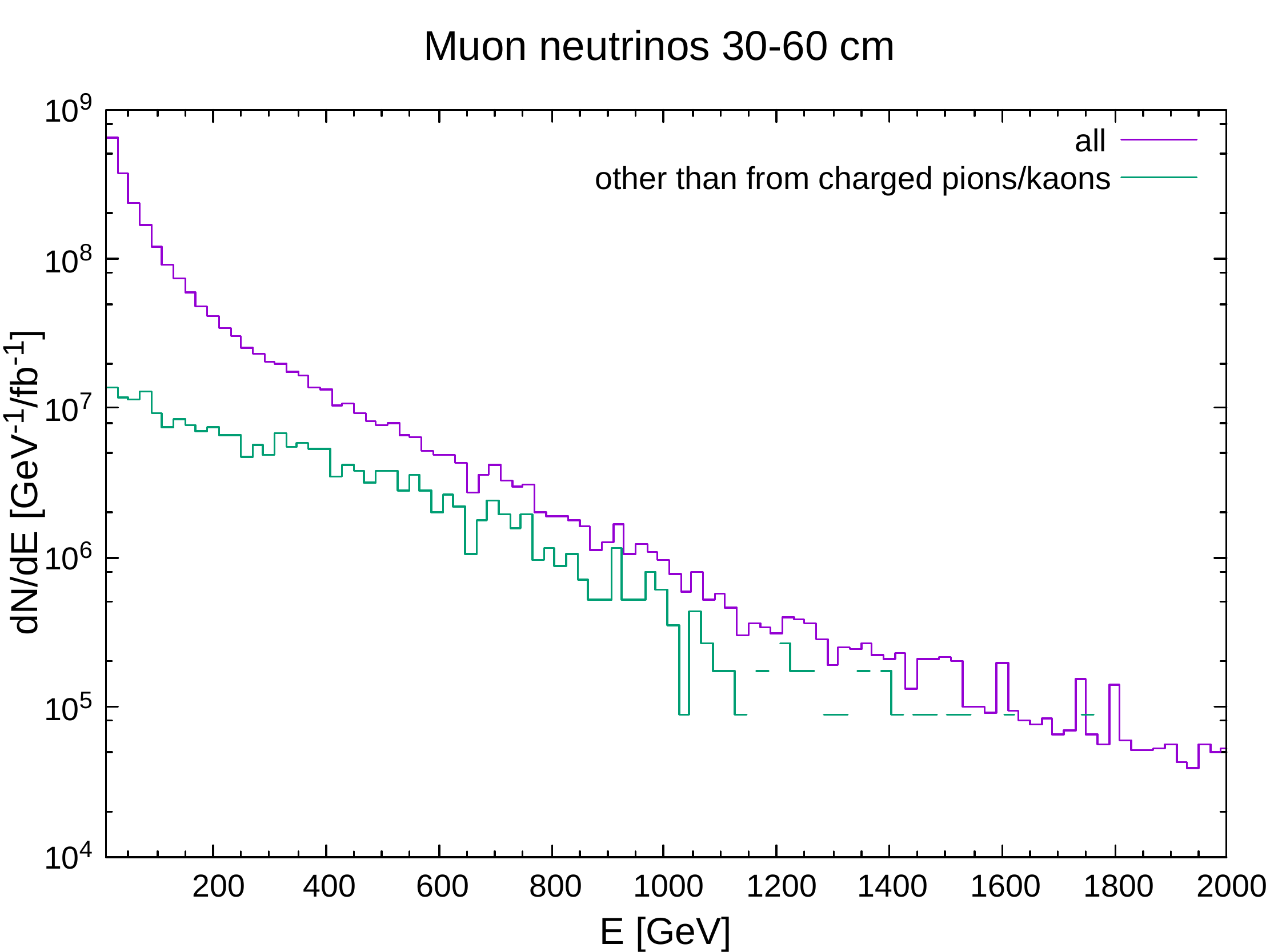}
    \includegraphics[width=0.55\textwidth] {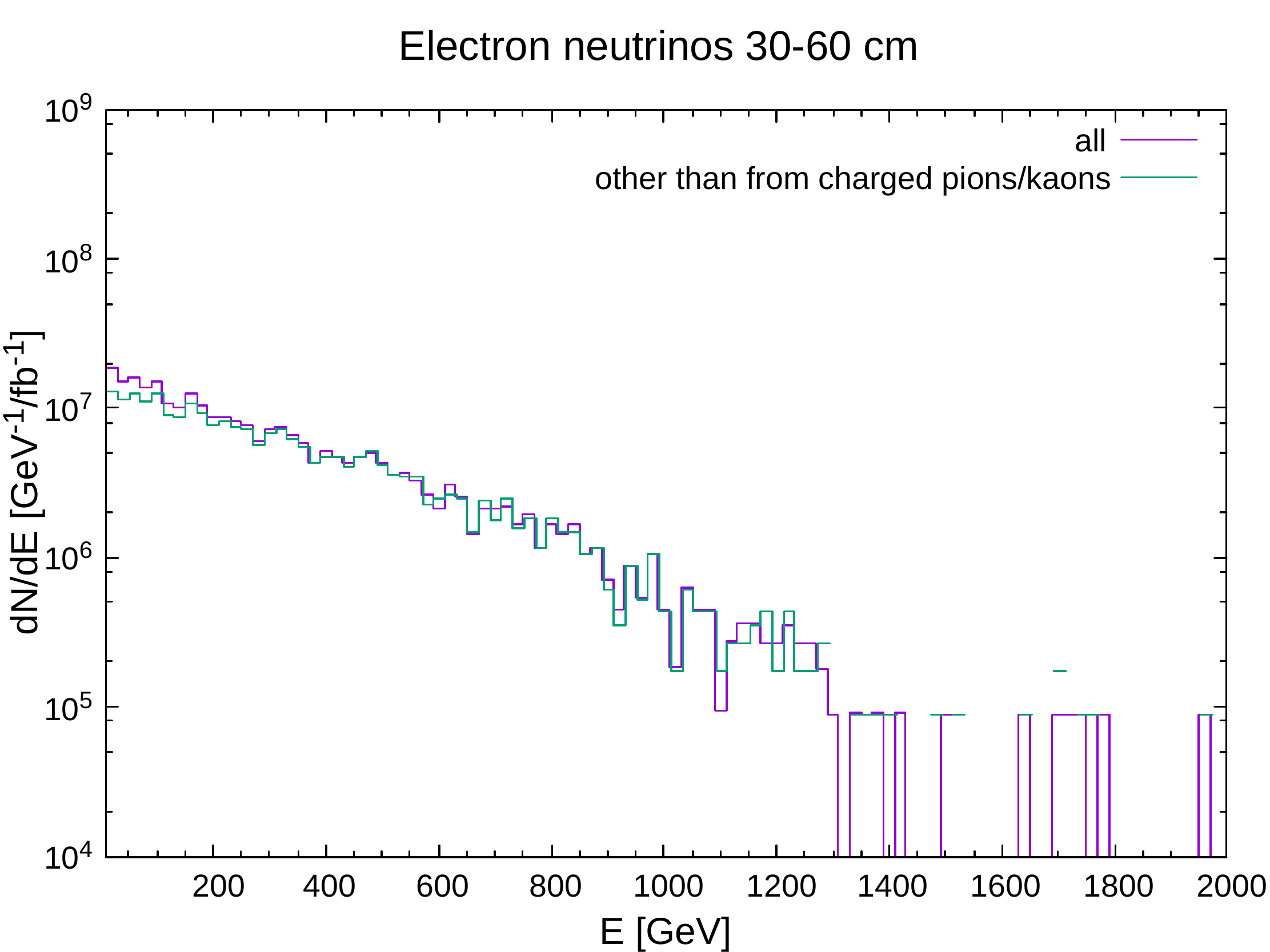}
   \includegraphics[width=0.6\textwidth] {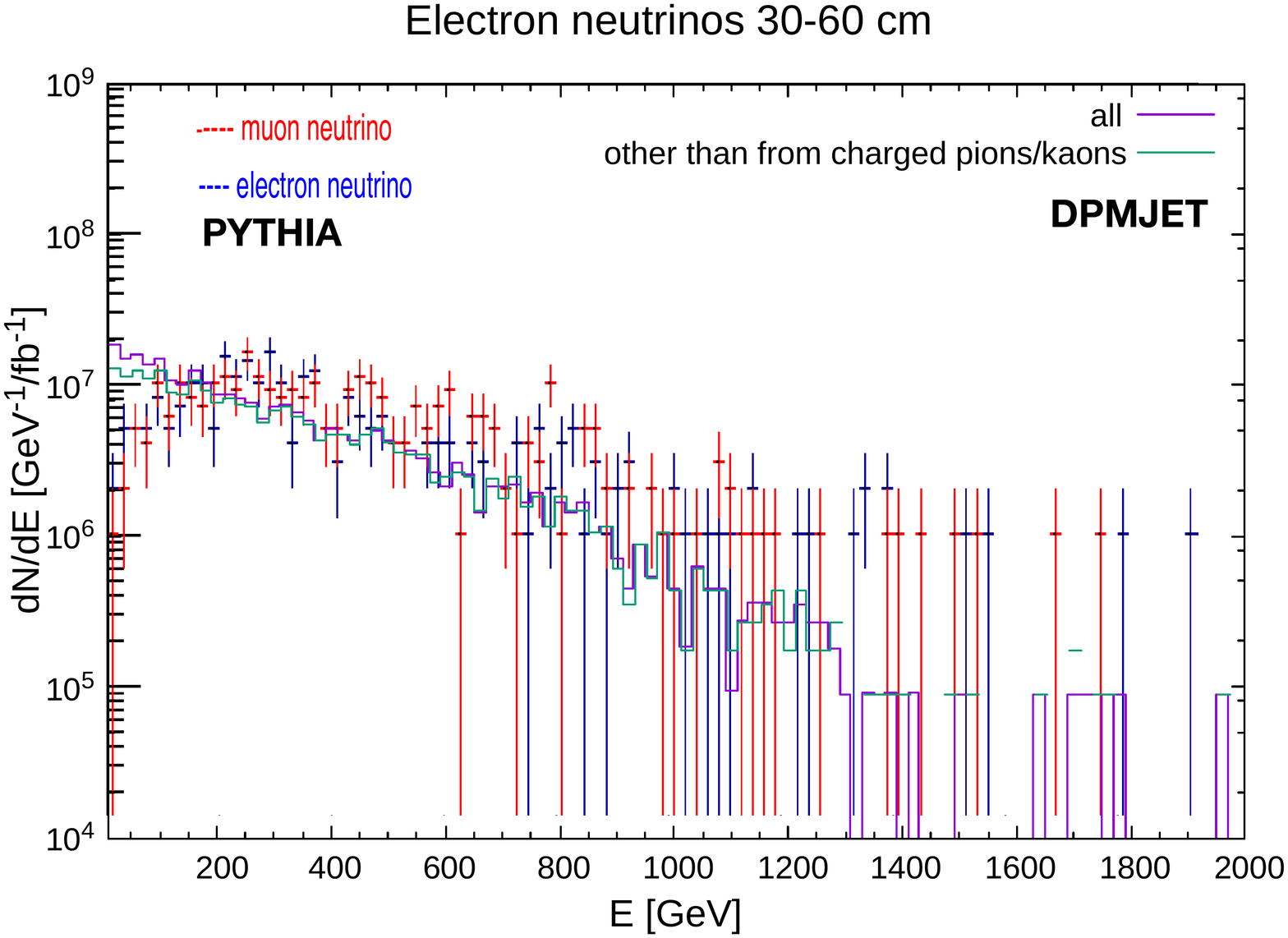}
\caption[B1B2 architecture] {Predicted fluxes of neutrinos for the radial distance $30<R<60$~cm from the beam axis in TI18 ($7.4<\eta<8.1$).
Top: muon neutrinos from DPMJET (Dual Parton Model, including charm) and LHC simulation. 
Middle: electron neutrinos from DPMJET and LHC simulation. 
Bottom: muon and electron neutrinos from PYTHIA  superimposed to DPMJET. 
 \label{fig:fluxes3060}}
\end{figure}

\begin{figure}[htbp]
  \centering
    \includegraphics[width=0.55\textwidth]{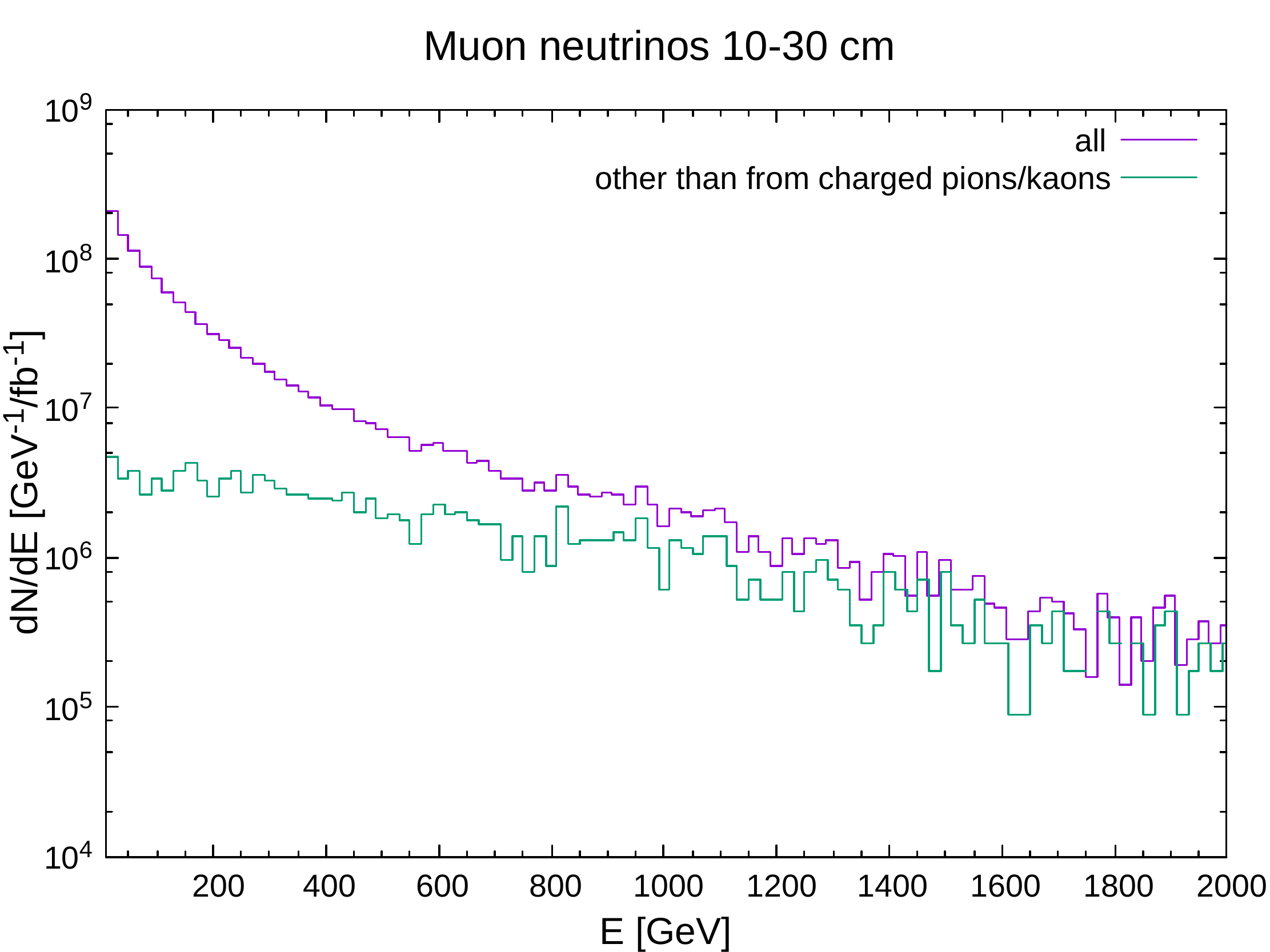}
    \includegraphics[width=0.55\textwidth] {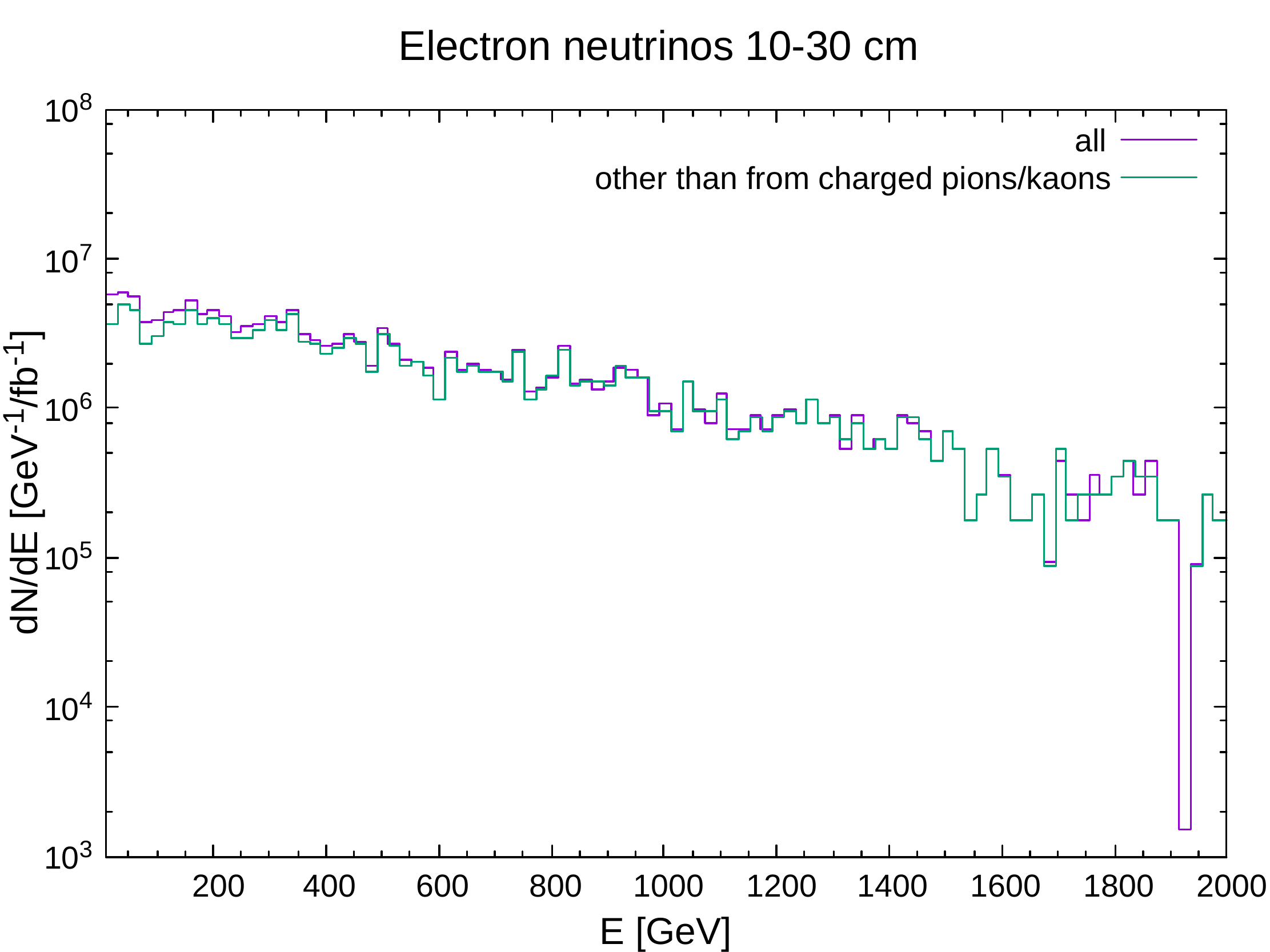}
     \includegraphics[width=0.6\textwidth] {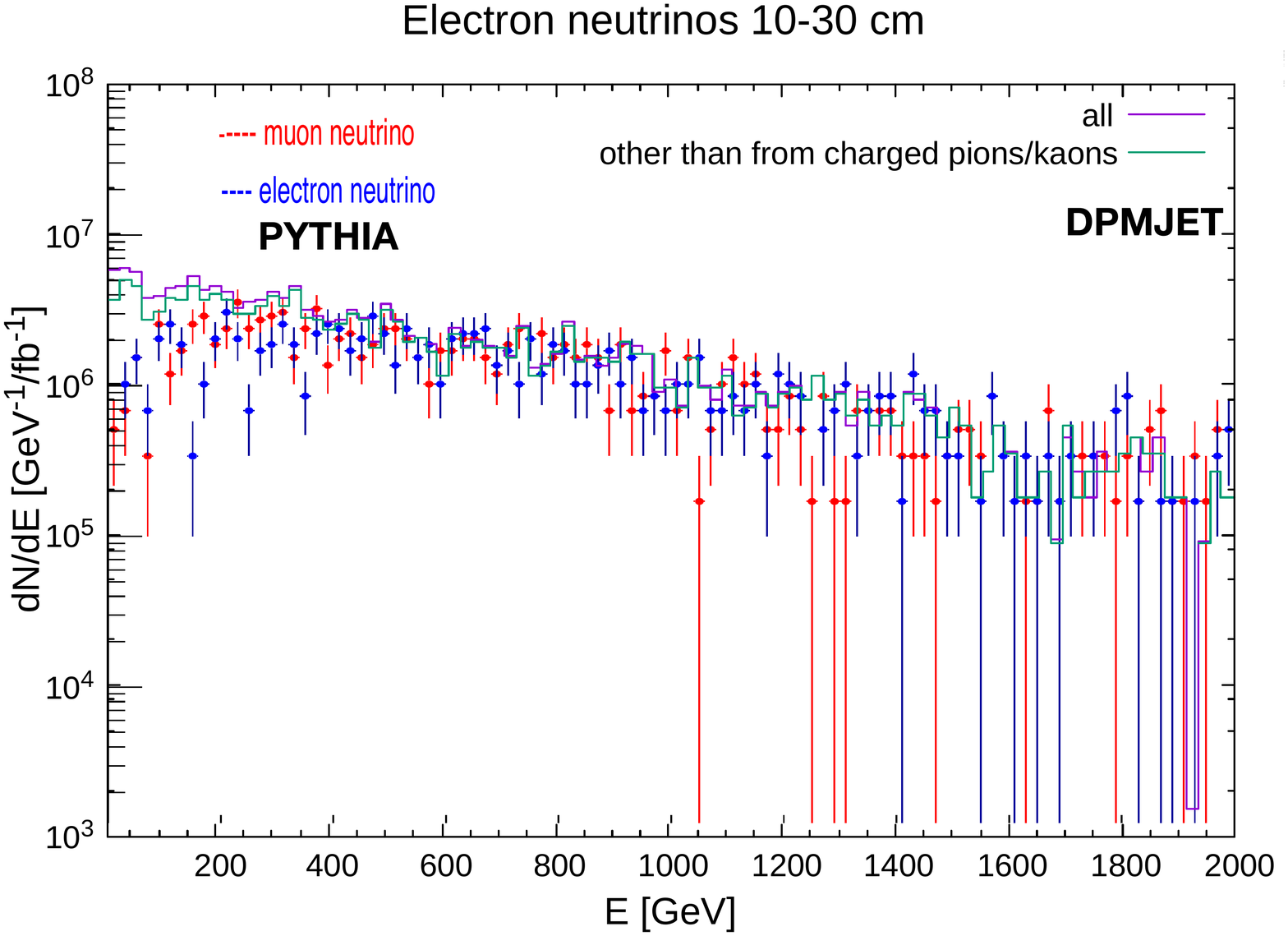}   
\caption[B1B2 architecture] {Predicted fluxes of neutrinos for the radial distance $10<R<30$~cm from the beam axis in TI18 ($8.0<\eta<9.2$).
Top: muon neutrinos from DPMJET (Dual Parton Model, including charm) and LHC simulation. 
Middle: electron neutrinos from DPMJET and LHC simulation. 
Bottom: muon and electron neutrinos from PYTHIA  superimposed to DPMJET. 
 \label{fig:fluxes1030}}
\end{figure}

In Figure~\ref{fig:fluxes3060}
neutrino energy distributions are plotted for the radial distance $30<R<60$~cm from the beam axis in TI18, corresponding to $7.4<\eta<8.1$, separately according to whether neutrinos originated from pion/kaon decays or not. The plots are consistent with the expectations that electron neutrinos and high energy muon neutrinos do not come from pion/kaon decays.
In Figure~\ref{fig:fluxes3060} the flux  of muon and electron neutrinos from charm production in LHC pp collisions
predicted with PYTHIA  is also shown; 
the absolute normalisatioin of the calculated flux is in excellent agreement with the DPMJET $\nu_{e}$ prediction, and it confirms
that electron neutrinos originate from charm.
In addition 
the agreement between the shapes of the distributions  is impressive when considering the different approach in  p$_T$ generation between PYTHIA
(heavy quark c and b production) and DPMJET (pp minimum bias).
In Figure~\ref{fig:fluxes1030}
the study is repeated for the radial distance $10<R<30$~cm from the beam axis in TI18, corresponding to $8.0<\eta<9.2$, 
with similar conclusions.

\begin{figure}[htbp]
 \centering
    \includegraphics[width=0.6\textwidth]{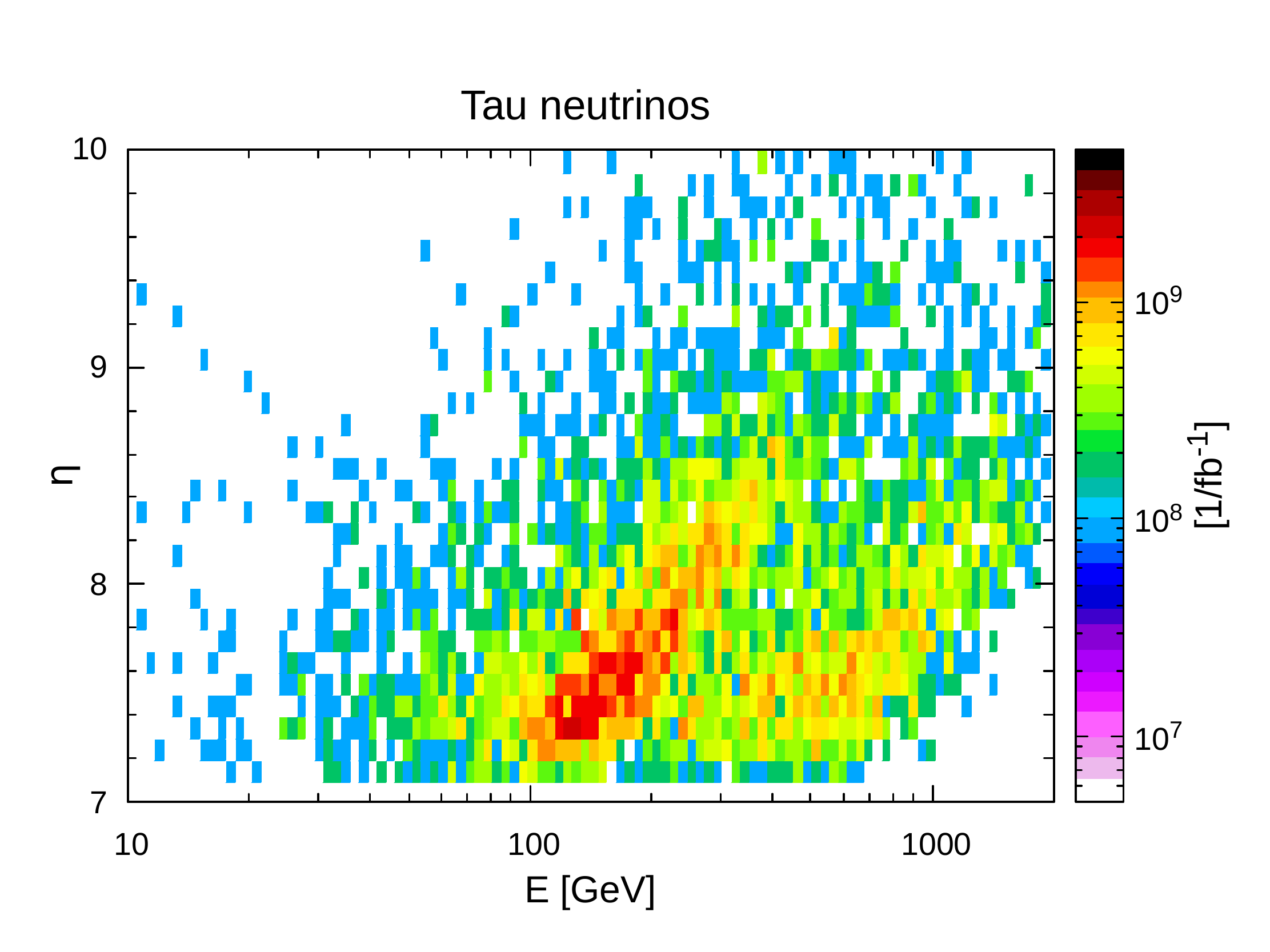}
    \includegraphics[width=0.52\textwidth] {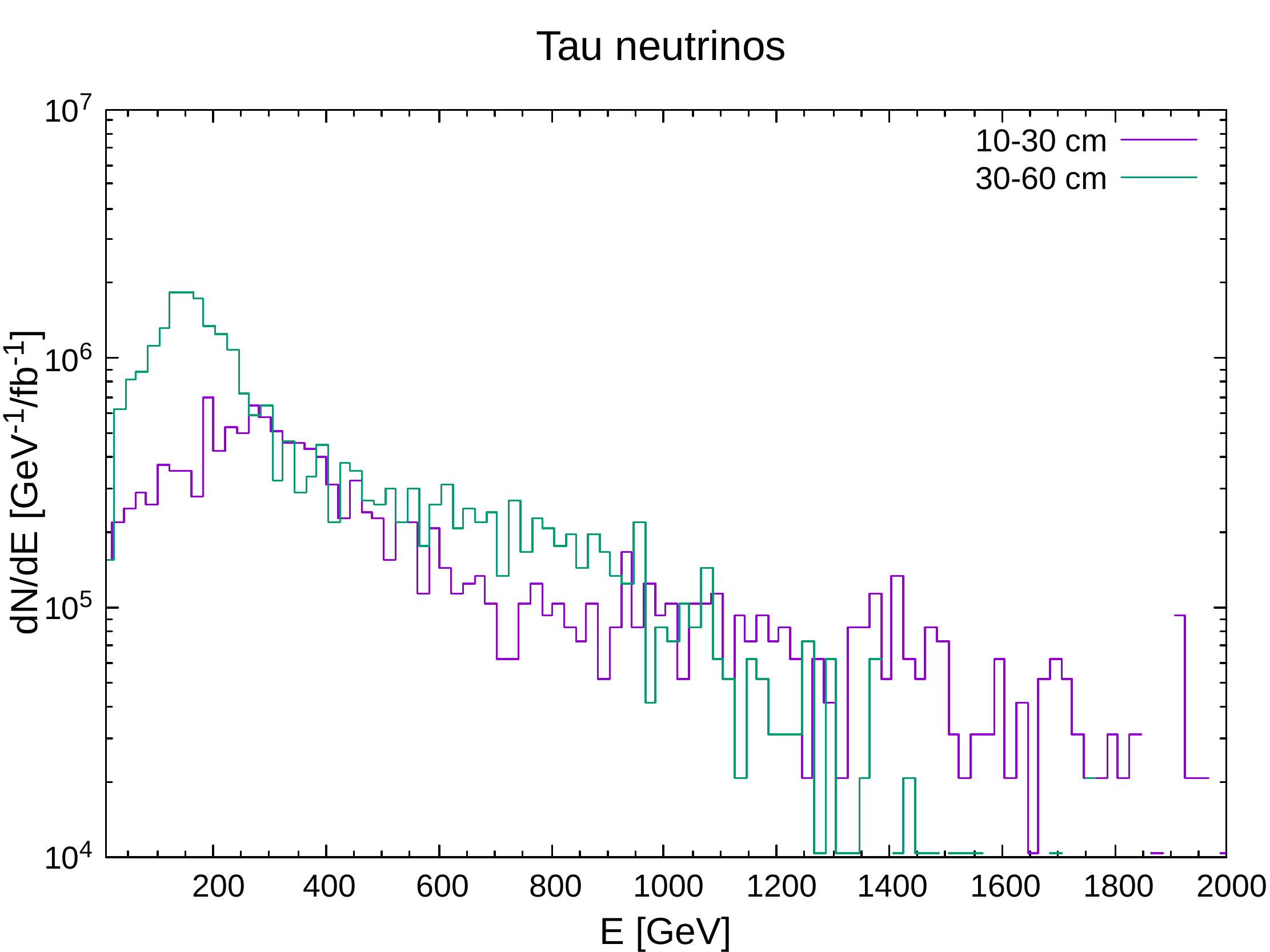}
    \includegraphics[width=0.6\textwidth] {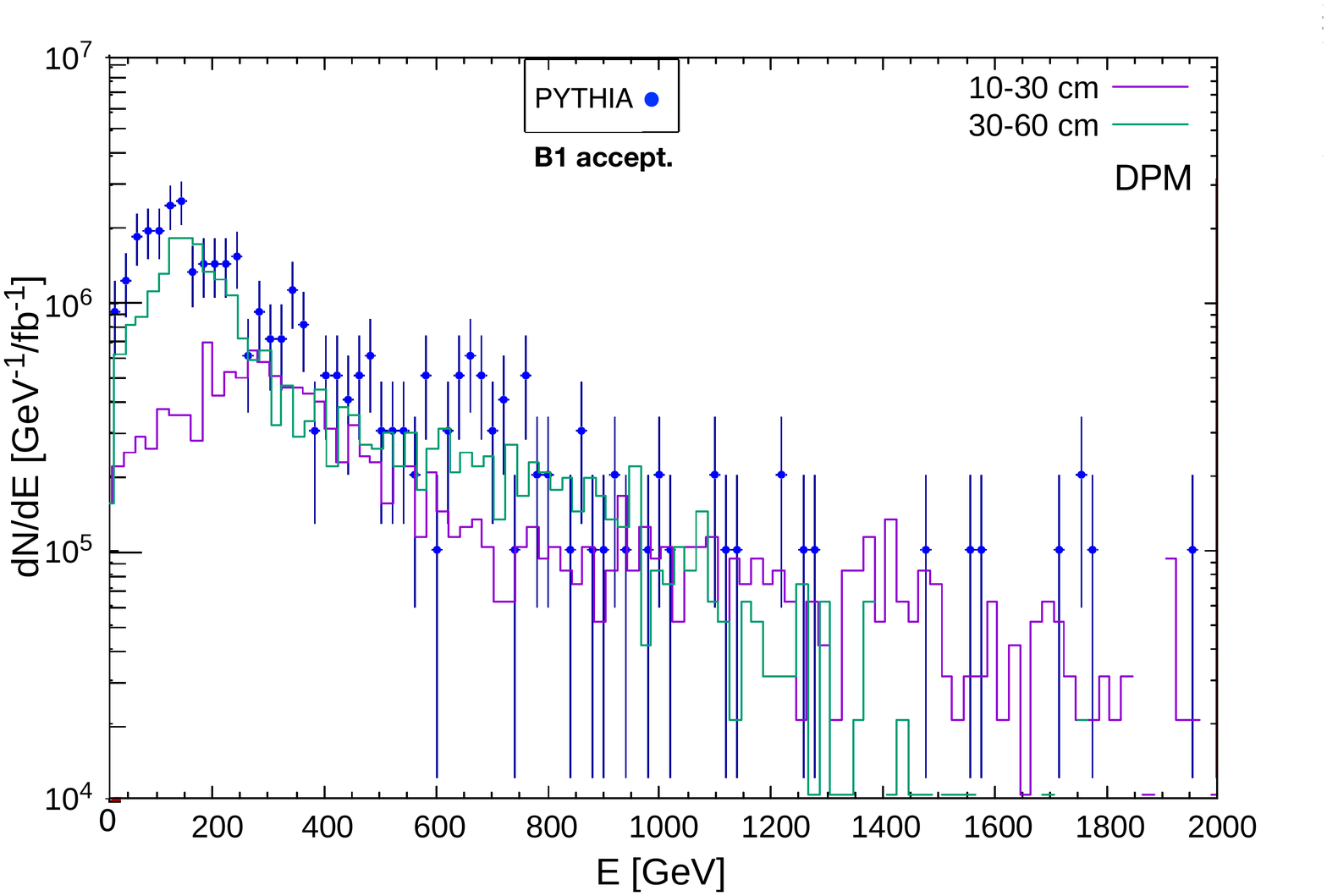}
\caption[B1B2 architecture] {Predicted fluxes of tau neutrinos for the radial distances $10<R<30$~cm ($8.0<\eta<9.2$) and $30<R<60$~cm ($7.4<\eta<8.1$) from the beam axis in TI18.
Top: Scatter plots of tau neutrino pseudorapidity $\eta$ versus energy. 
Events generated with FLUKA 
using the embedded DPMJET event generator, and the LHC optics simulation. 
Middle: energy spectra of tau neutrinos from DPMJET and LHC simulation. 
Bottom: PYTHIA (charm production) tau neutrino energy distribution superimposed to DPMJET. 
 \label{fig:fluxes_nutau}}
\end{figure}

Tau neutrinos are produced  in $D_{s}\rightarrow\tau\nu_{\tau}$ decays and in the subsequent $\tau$ decays.  
Since the chosen $\eta$ range  has shown high sensitivity to charm decays, we do expect this slightly off-beam-axis configuration to be favourable for tau neutrinos.
Figure~\ref{fig:fluxes_nutau} 
shows  the $\eta$ versus energy scatter plot and  the energy distribution for tau neutrinos in pp collision events generated with DPMJET/FLUKA and with PYTHIA.  They confirm the expectations. The flux calculations are normalised to 1~fb$^{-1}$. 
Note that the two intensity bands in the $\eta$ vs ln(E) scatter plot are predicted from kinematics: 
neutrinos from $D_{s}$ decays have little p$_T$ since the mass difference $M_{D_{s}}-M_{\tau}$ is small (191 MeV) and they populate the band at lower E; the other band is due to neutrinos from $\tau$ decays which do have a larger average  p$_T$ ($\approx$592 MeV in three body decays).

In summary, 
the simulations performed using both PYTHIA and DPMJET 
agree in showing that a detector acceptance slightly off beam axis,  in the  
 $7.4<\eta<9.2$ range, is favorable for looking at high energy neutrinos from charm production in pp collisions and consequently for observing tau neutrinos. 
The state-of-the-art LHC optics emulation with FLUKA  shows that this physics potential 
is not spoiled 
by the flux of muon neutrinos from
pion and kaon decays which is concentrated at lower energies.

\newpage

\begin{figure}[hptb]
  \centering
    \includegraphics[width=0.6\textwidth]{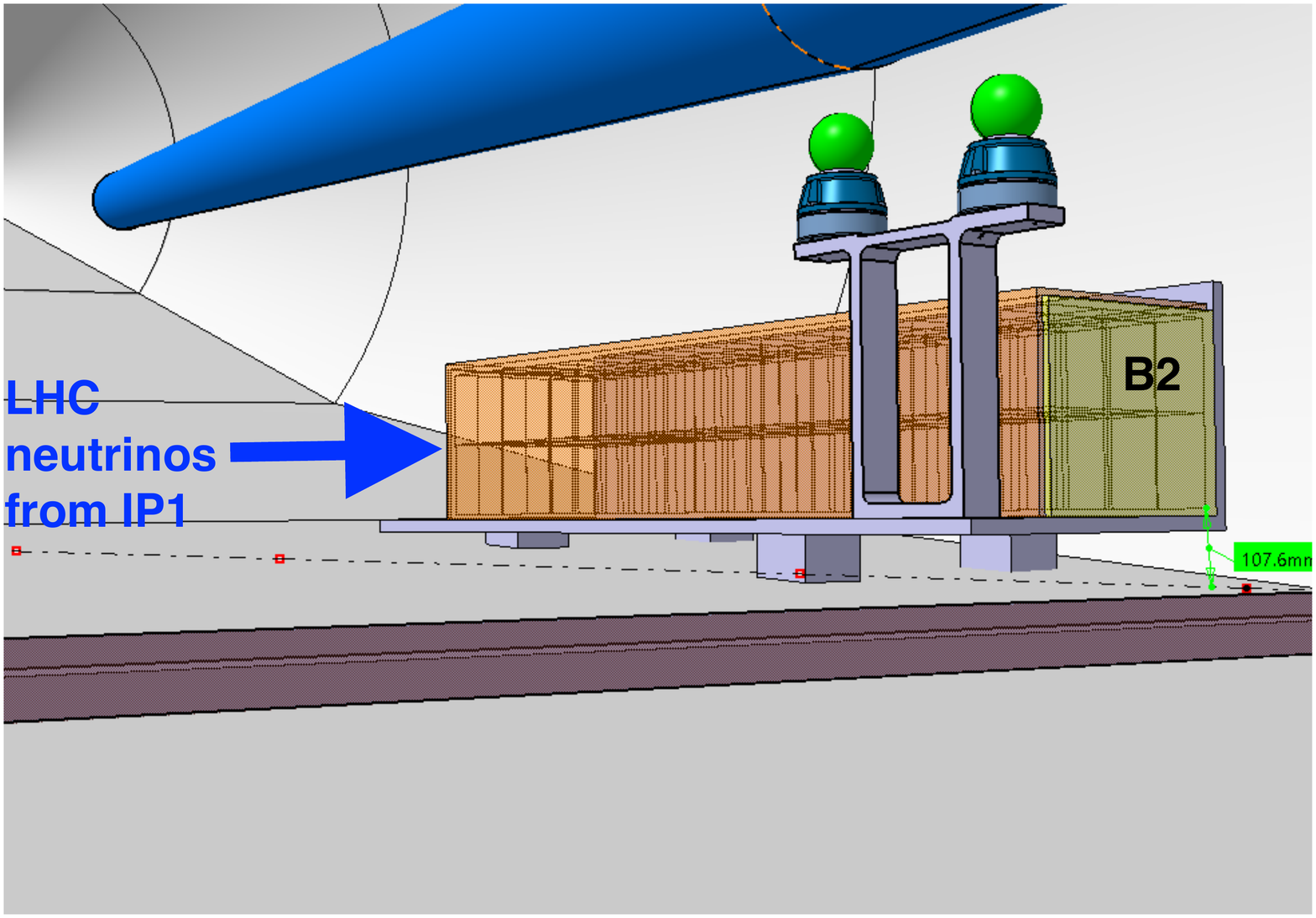}
    \includegraphics[width=0.45\textwidth] {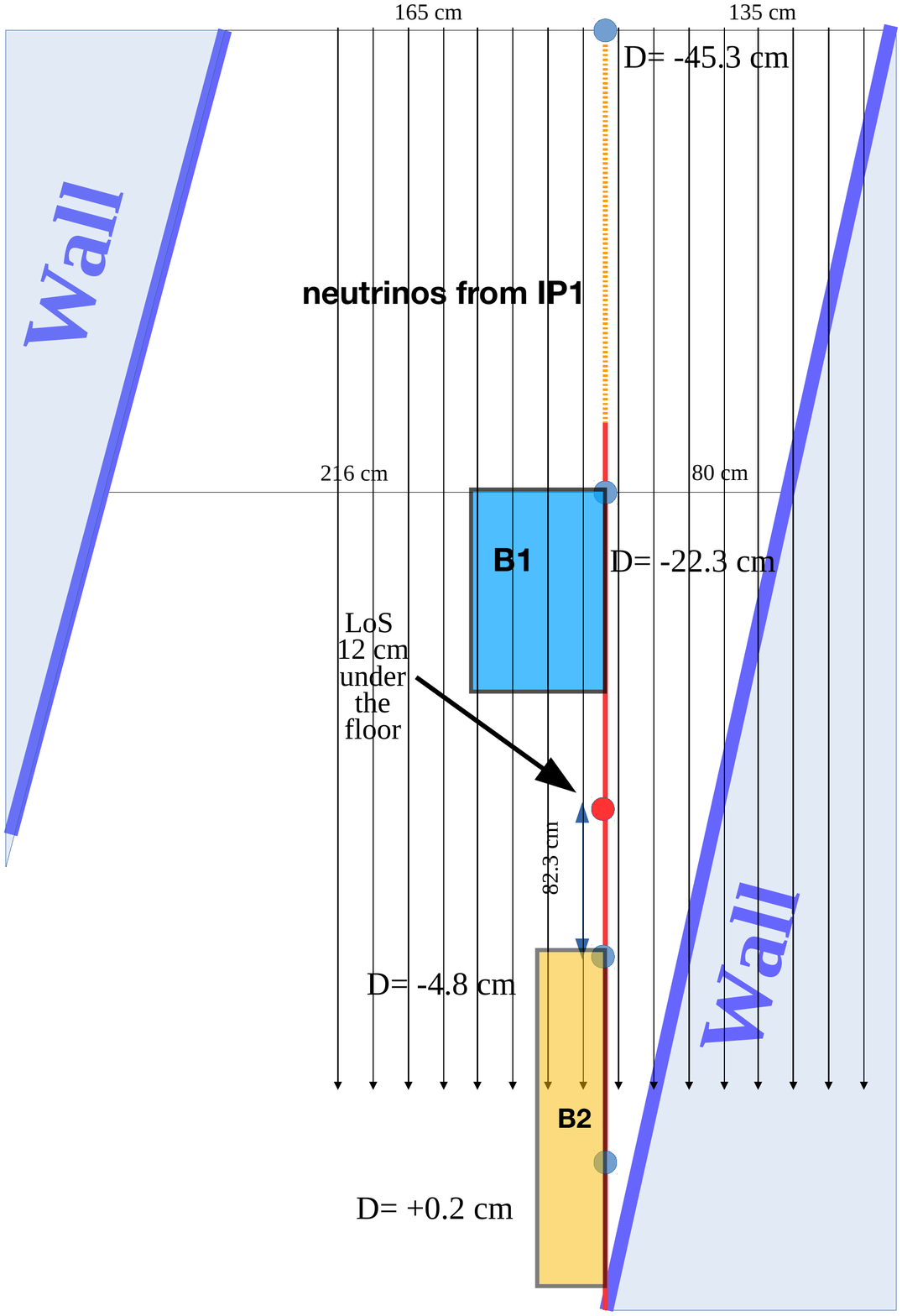}
    \includegraphics[width=0.45\textwidth] {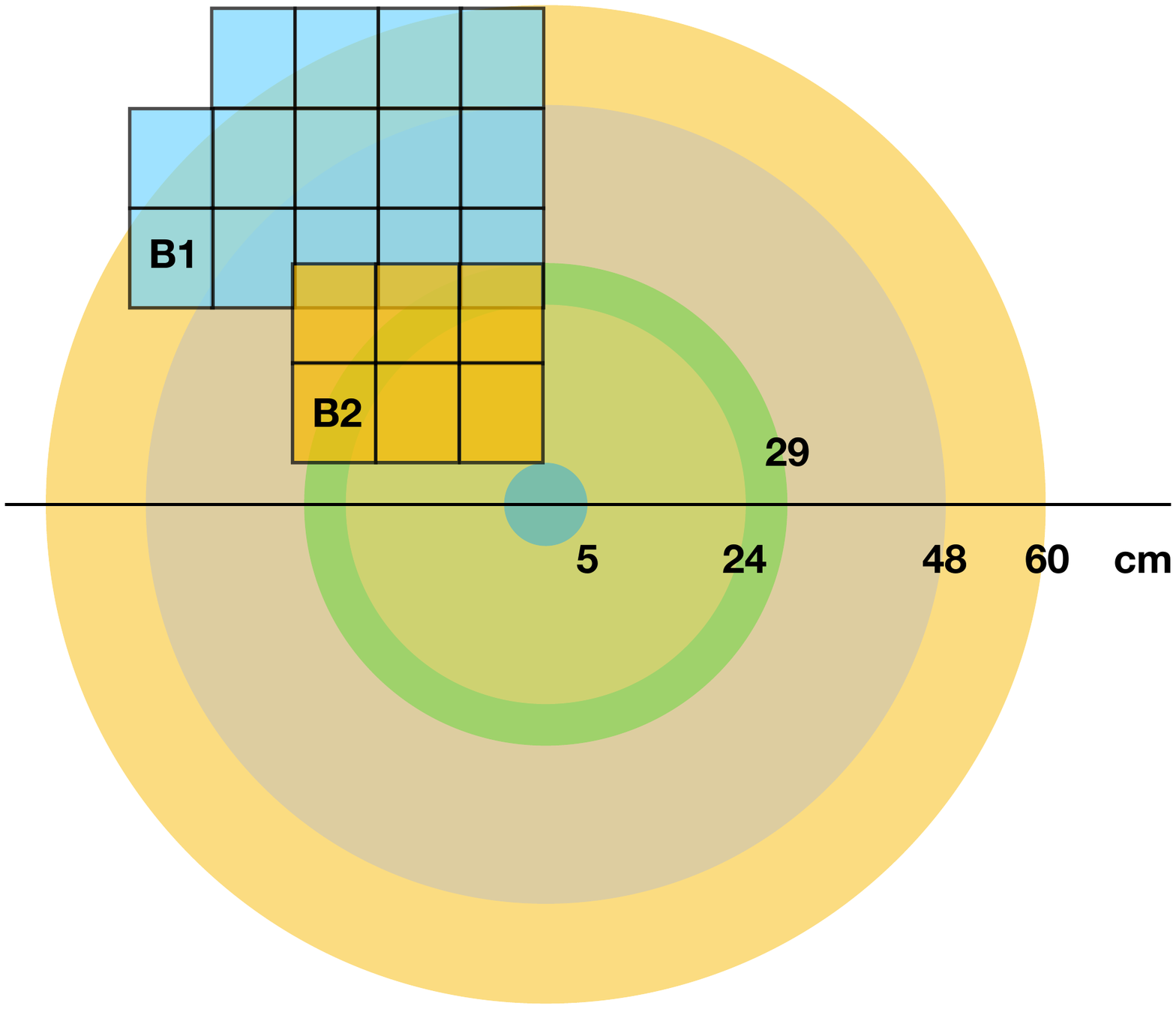}
\caption[B1B2 architecture] {A possible architecture of the B2 and B1  detectors;
the numbering refers to the distance to the IP: B1 is closer. Top: 3D view of B2
\cite{CERN-EN}. 
Bottom left: top view of B1 and B2; D is the height of the the LoS with respect to the TI18 cavern floor, measured in different points. Bottom right: section view of B1 and B2 showing their radial distance in cm from the beam axis, which is taken as 57.6~mm above the LoS.
 \label{fig:XSEN_B1B2}}
\end{figure}

\section{Predicted Numbers of Events}

In the following we will consider two independent detectors as shown in 
Figure~\ref{fig:XSEN_B1B2}, subtending two different $\eta$ ranges as discussed in 
Section
~\ref{sec:ExpGen}.
The studies are based on large samples of simulated pp collisions: 10~M PYTHIA events of heavy hadron (c and b) production, 50~M DPMJET/FLUKA minimum bias events (extended to 400~M for the tau neutrino predictions).

 A 3D view of a compact detector ( B2 )
that subtends the pseudorapidity range $8.0<\eta<9.2$ is shown 
in Figure~\ref{fig:XSEN_B1B2} top. 
It consists of 108 OPERA bricks (3x2x18, ~1.4 m long) and weighs 0.9~tons.  
The B2 detector is positioned  108 mm above the LoS:
 the beam axis 
in TI18
is centred high above the LoS by 57.6~mm, 
since the proton beams in IP1 cross vertically with a half-angle of 120~microradians.
We expect that during Run~3 the crossing half-angles in IP1 will change in the same range as during 2018.
That means that 
during fills the angles will go from 160 down to 120~microradians, 
corresponding to 76.8 mm and 57.6 mm of LoS displacement at TI18 respectively. 
A second detector ( B1 ) is placed further uphill in TI18, towards IP1. 
Figure~\ref{fig:XSEN_B1B2} shows an implementation with 168 OPERA bricks (14 bricks thick). 
Space allows for  doubling the azimuthal acceptance. 
It has acceptance for $7.4<\eta<8.1$ and weighs 1.4 tons.

Inversion from upwards to downwards of the LHC beam crossing at the ATLAS IP during Run~3 will modify the detector acceptance. It will shift to $7.7<\eta<8.4$ in B2 and to $7.2<\eta<7.8$ in B1. This slight change does not substantially modify the nature of the neutrino flux 
already discussed in Section
~\ref{sec:NuFlux}. 
The event rate will reduce by $\approx$20\%. 
It is foreseen that about half of the luminosity will be delivered in each configuration. Run~3 was extended to include 2024 and it  is expected to integrate more than 150~fb$^{-1}$ and possibly as much as 200-250~fb$^{-1}$. In the following, the calculated event numbers are relative to 
150~fb$^{-1}$ with upward beam crossing.

\begin{figure} [b]
\centering
\includegraphics[width=0.7\textwidth]{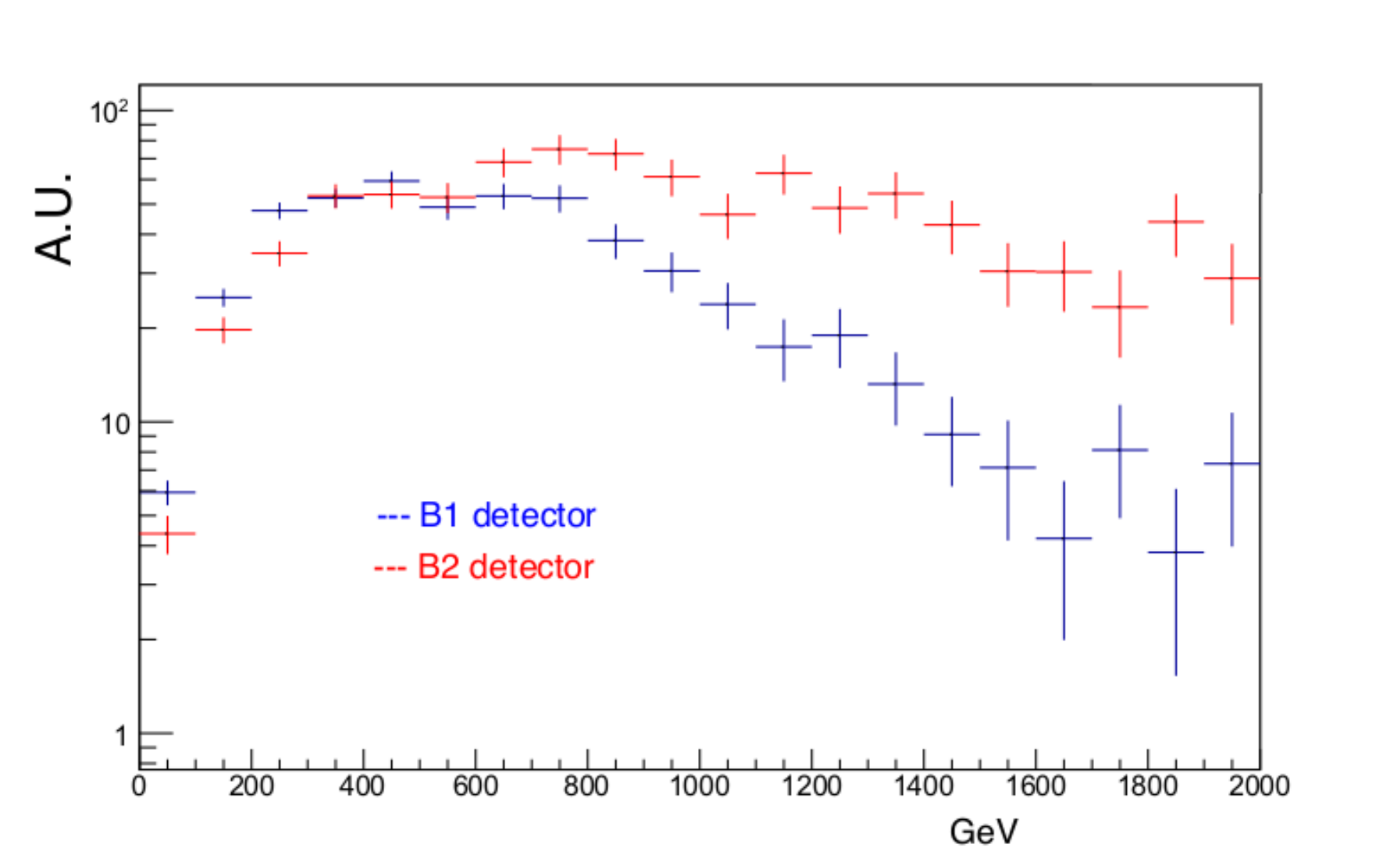}
\caption[spectrum] {PYTHIA predicted energy spectra of neutrinos from charm and bottom decays interacting in the B1 and B2 detectors.
\label{fig:DNDE_B2_B1}}
\end{figure}

Simulations of proton-proton interactions at 13~TeV were performed with PYTHIA.
An event by event weight was applied for properly taking into account the $\nu$N CC interaction cross section dependence on energy and on neutrino flavour
\cite{nutauN, PDG}. 
Table~\ref{tab:eventrate}  
summarizes the expected numbers and characteristics of the CC events for the detector configuration shown in 
Figure~\ref{fig:XSEN_B1B2}.
The predicted energy spectra of neutrino CC events in both the B1 and B2 detectors are shown in Figure~\ref{fig:DNDE_B2_B1}.
The spectrum is biased towards high energies, since the $\nu$N cross section grows rapidly with neutrino energy.
\begin{table} [b]
\begin{center}
\topcaption{ Neutrino CC event rate expectations for a detector configuration as shown in 
Figure~\ref{fig:XSEN_B1B2}
for an LHC luminosity of 150~fb$^{-1}$. 
B1 (1.4 tons) and B2 (0.9 tons)  acceptances subtend 
 $7.4<\eta<8.1$ and $8.0<\eta<9.2$
respectively. 
Neutrino flux from c and b decays calculated with PYTHIA.}
\label{tab:eventrate}
\begin{tabular} {lrrrrr} \hline 
& & & & \\
  &B1  &B2 &B1+B2  &2xB1+B2  \\  
& & & & \\ \hline
integral $\nu$ fluence   &4.6$\times$10$^{11}$ &3.4$\times$10$^{11}$   &0.8$\times$10$^{12}$  &1.3$\times$10$^{12}$    \\
all  $\nu$ events  &490 &852   &1342  &1832    \\
tau flavour $\nu$ events   &26 &25   &51  &77  \\ 
$\eta$ range &7.4-8.1 &8.0-9.2 & & \\
average E$_{\nu}$ (RMS) GeV  &700(400) &1200(600) & &  \\
& & & & \\ \hline
\end{tabular}
\end{center}
\end{table}
\begin{table} [t]
\begin{center}
\topcaption{ Expectations for neutrino CC  event rate and average energy,
 calculated with DPMJET/FLUKA event generator and
the simulation of the LHC magnetic optics,
for an LHC luminosity of 150~fb$^{-1}$. Detector
B1 (1.4 tons) and B2 (0.9 tons)  acceptances subtend 
 $7.4<\eta<8.1$ and $8.0<\eta<9.2$
respectively.  }
\label{tab:DPMJETeventrate}
\begin{tabular} {lrrrrrrr} \hline 
& & & & & &\\
  & &B1 & & & &B2 &\\ 
   &all & &exclude  & &all & &exclude\\ 
   & & &ch. $\pi$,K  & & & &ch. $\pi$,K\\  \hline
& & & & & &\\
$\mu$ flavour $\nu$ events  &929 & &267 &   &1516 &  &538    \\
$e$ flavour $\nu$ events  &302 & &292 &   &616 &  &598    \\
$\tau$ flavour $\nu$ events   &16 & &16  & &28 & &28  \\ 
all $\nu$ events  &1247 & &575 &   &2160 &  &1164    \\
& & & & & &\\
average E$_{\nu}$ (RMS) GeV  & & & &  \\
$\mu$ flavour $\nu$ events  &385(400) & &535(315) & &740(685) & &1145(710)  \\
$e$ flavour $\nu$ events  & & &560(350) & & & &1165(715)  \\
$\tau$ flavour $\nu$ events  & & &630(370) & & & &1095(720)  \\
& & & & \\ \hline
\end{tabular}
\end{center}
\end{table}

\begin{figure}[b]
  \centering
    \includegraphics[width=0.6\textwidth]{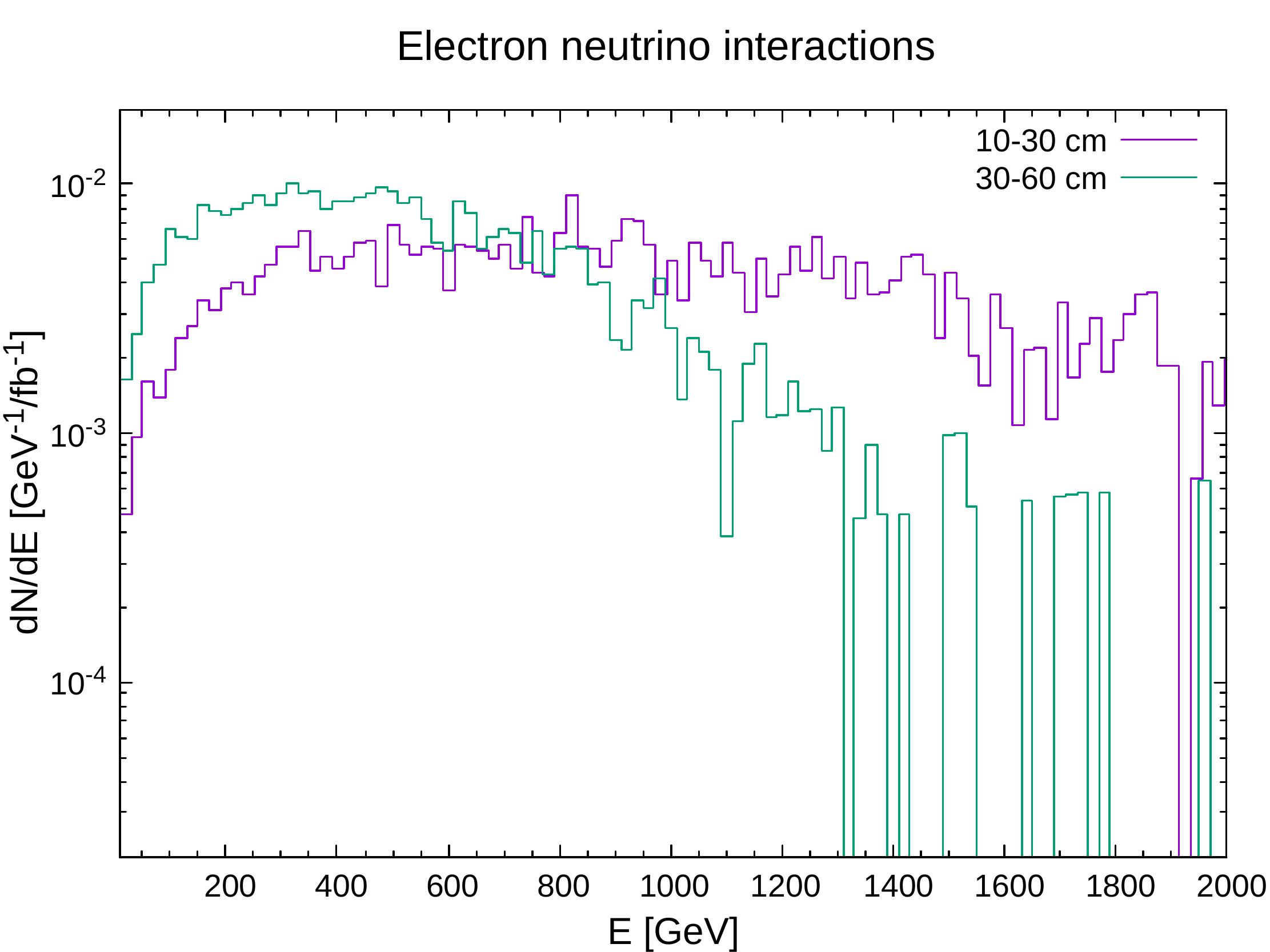}
    \includegraphics[width=0.6\textwidth] {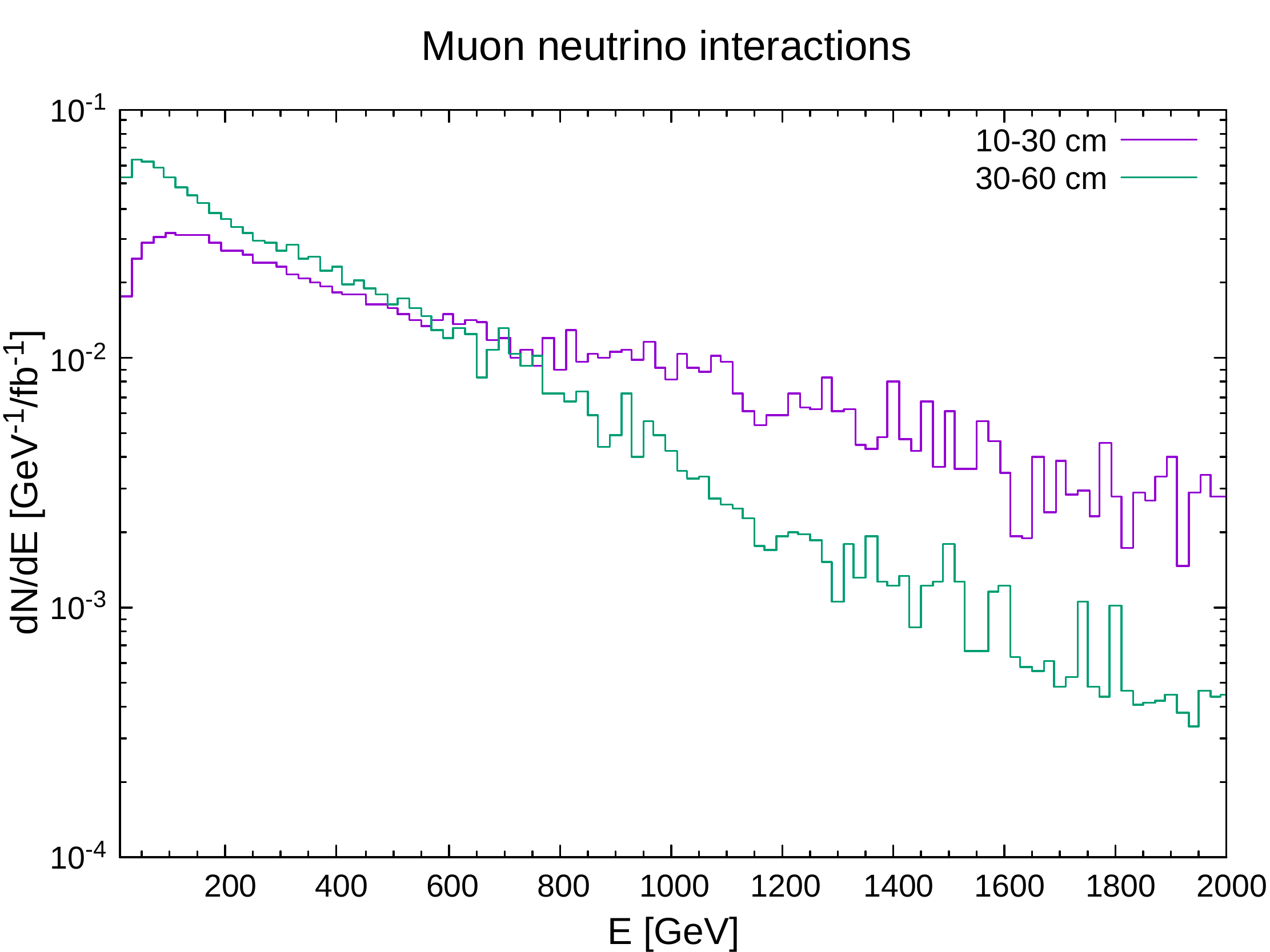}
\caption[B1B2 architecture] {
DPMJET/FLUKA predicted energy spectra of neutrinos  interacting in the B1 (30-60 cm radially out from beam axis) and B2 (10-30 cm radially out) detectors.
 \label{fig:nuint_spectra}}
\end{figure}

Calculation with PYTHIA accounts for the contribution from decays of heavy quarks (c and b). The additional contribution of pion and kaon decays was studied  by the CERN FLUKA team of EN-STI
using DPMJET.
Charged pion and kaons are transported along the LHC straight section and through the LHC magnetic optics until they decay.
The DPMJET minimum bias event generator also includes charm.
The same detector response function is used as for PYTHIA calculations.
The results are shown in Table~\ref{tab:DPMJETeventrate}.  

The predictions for electron and tau neutrino rates of PYTHIA and DPMJET/FLUKA are in good agreement, and their spectra are consistent. 
The $\nu_{e}$ and $\nu_{\tau}$ events in the B1 and B2 detectors probe charm production and neutrino physics in two rather independent energy bins.
For muon neutrinos, PYTHIA and DPMJET compare well,
when neutrinos from pion and kaon decays, which populate the  lower energy part of the spectrum in DPMJET, are excluded. 
Figure~\ref{fig:nuint_spectra}
shows the DPMJET predicted energy spectra of 
electron and muon neutrino CC events in the two pseudorapidity ranges under consideration.

\begin{figure}[t]
  \centering
    \includegraphics[width=0.55\textwidth]{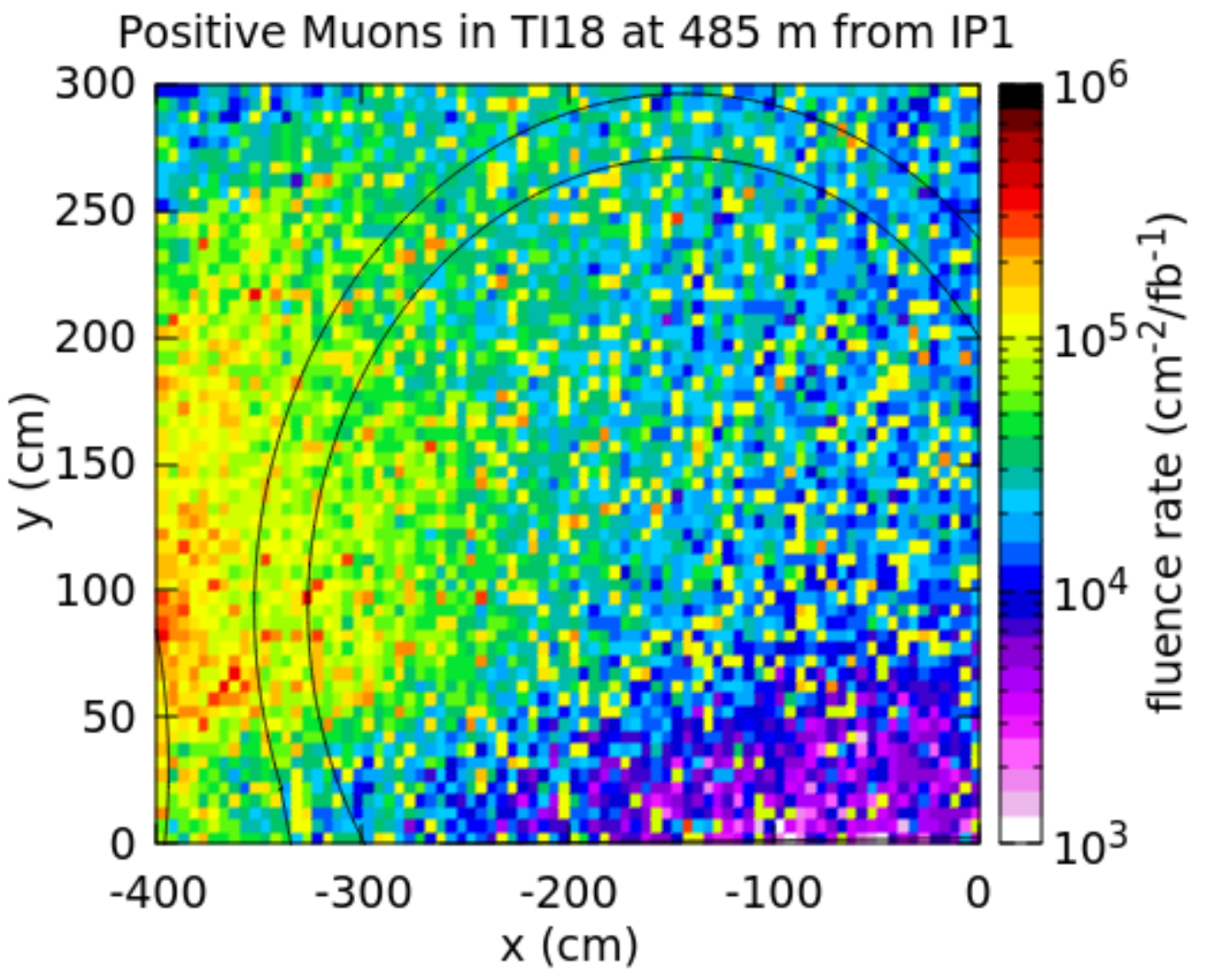}
    \includegraphics[width=0.55\textwidth] {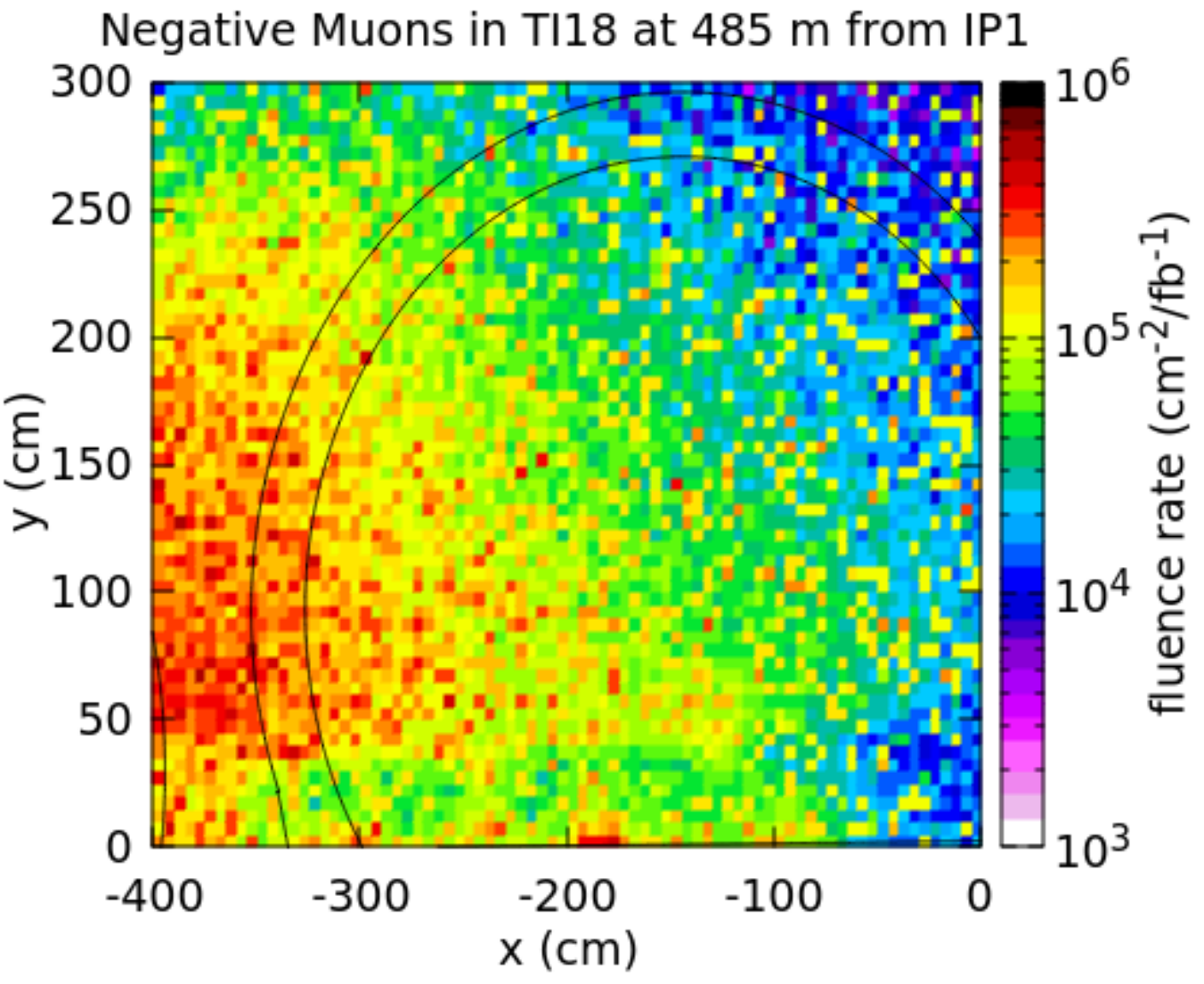}
\caption[B1B2 architecture] {
Muon flux in TI18 calculated with DPMJET/FLUKA  and
the simulation of the LHC magnetic optics.
The pictures show a slice of the TI18 cavern, 485~m from IP1.  The origin (0,0) is set on the LoS. Dimensions are centimeters. The black lines delimit the tunnel wall.
 \label{fig:muon_fluxinTI18}}
\end{figure}

The detailed calculations on pion and kaon transportation along the LHC optics and their decay also provide an estimate of the muon flux in TI18 ( Figure~\ref{fig:muon_fluxinTI18}). 
In the region occupied by B1 and B2  it is expected that
10$^{3}$-10$^{4}$~muons/cm$^{2}$ in 1~fb$^{-1}$ of LHC luminosity will be recorded:
they provide an excellent tool for aligning the detector emulsion layers.

\clearpage 
\newpage
\section{Summary and Outlook}

This work complements a previous paper 
\cite{XSEN1}
that discussed the phyics potential of a neutrino experiment at the LHC, and 
identified  a favorable site.
The requirements were: high flux of TeV neutrinos; sizeable contribution of tau neutrinos; low machine backgrounds. 
It suggested that a detector could be positioned in the
cavern of the TI18 (or TI12) tunnel  that intercepts the LHC  after the beginning of the arc,
$\approx$480 m from the ATLAS IP.

In this paper, global features and physics reach of such an experiment located in TI18 are investigated.
For simplicity the studies are based on a detector that introduces minimal perturbation in the LHC: it consists of layers of lead and emulsions, no magnetic field, no active electronics components.
However, the results apply to any detectors with similar mass and similar $\eta$ acceptance.
Our conceptual detector  has a total mass in the range 2.3-3.7~tons, covers the angular acceptance $7.4<\eta<9.2$ and takes data during  the LHC Run~3, integrating a luminosity of 150~fb$^{-1}$.

Because the neutrino-nucleon interaction cross section grows with energy,  a low target mass
can be used, 
and consequently the spectrum of observed neutrinos is biased towards higher energies.

The detector position, which is slightly off the LHC beam Line-of-Sight, is optimised for neutrinos
originating at the IP.
In this $\eta$ range most high energy neutrinos are
from  c ($\approx$92\%) and b ($\approx$8\%) decays; about 5\% of the neutrinos are of the tau flavour, from $D_{s}\rightarrow\tau\nu_{\tau}$ decays and subsequent $\tau$ decays.
Additional muon neutrinos from pion and kaon decays are
mostly
at lower energies.

Simulations are performed using PYTHIA 
\cite{Pythia}
and DPMJET/FLUKA
\cite{Fluka1, Fluka2, DPMJET, LHCsim}.
The FLUKA set-up used by the CERN EN-STI team allows for tracking charged pions and kaons along the LHC machine until they decay.  
The integral neutrino fluence on the detector is estimated to exceed 10$^{12}$~neutrinos in 
150~fb$^{-1}$.  

The $\eta$ acceptance is split into two regions, $7.4<\eta<8.1$ (B1) and $8.0<\eta<9.2$ (B2),
which have slightly different kinematics constraints and neutrino flux compositions, 
and are independently studied. 
Electron neutrinos are shown to originate at the IP, mainly from charm; their behaviour is consistent with kinematics predictions.
Their flux and spectra are in excellent agreement between PYTHIA and DPMJET, in both $\eta$ ranges.
Similar results hold for tau neutrinos.
Muon neutrinos at high energy  are originating at the IP from charm and their spectrum is consistent with that of electron neutrinos, while at low energy the muon neutrino spectrum is dominated by pion and kaon decays.

Assuming masses of 1.4~tons and 0.9~tons for the B1 and the B2 detector respectively, consistent with the available space in the TI18 cavern,  
it is shown that the experiment can collect a few thousand high energy electron and muon neutrino CC~interactions,
and over 50 tau neutrino CC events,  in 150~fb$^{-1}$ of delivered LHC luminosity.
The average energies are $\approx$600~GeV in B1 and $\approx$1.1~TeV in B2 , 
with large RMS of $\approx$0.5~TeV.
These can be used for a first look at untested physics domains, as 
in charm production in proton-proton collisions  or in $\nu$N TeV interactions.
At HL-LHC such an experiment can collect ten times more events, but it will need active detectors
and adequate infrastructures, and it could inspire 
a new class of neutrino experiments at future very high energy colliders.  

\begin{acknowledgments}
We would like to thank  I.~Ajguirey, A.~Ball,  T.~Camporesi, A.~Dabrowski,  D.~Dattola, F.~Gasparini, V.~Klyukhin, M.~Komatsu. S.~Mallows, A.~Perrotta, V.~Tioukov, W.~Zeuner and the CERN EN/SMM-RME group  for help in various parts of this study, .
\end{acknowledgments}

\bibliography{auto_generated}   

\providecommand{\href}[2]{#2}\begingroup\raggedright\begin{thebibliography}{10}%
\makeatletter
\providecommand{\hrefCMSnoop }[0]{\@secondoftwo}%
\makeatother
\providecommand{\doi}{\texttt{doi:}\begingroup \urlstyle{tt}\Url}

\bibitem{XSEN1}
\hrefCMSnoop {}{N.~Beni {et~al.}, ``{P}hysics {P}otential of an {E}xperiment
  using {LHC} {N}eutrinos'',} \textit{ J. Phys. G: Nucl. Part. Phys.} \textbf{
  46} (2019) 115008,
  \href{http://dx.doi.org/10.1088/1361-6471/ab3f7c}{\doi{10.1088/1361-6471/ab3f7c}}.

\bibitem{PDG}
\hrefCMSnoop {}{{(Particle Data Group)}, M.~Tanabashi {et~al.}, ``{Review of
  Particle Physics }'',} \textit{ Phys. Rev.} \textbf{ D 98} (2018) 030001,
  \href{http://dx.doi.org/10.1103/PhysRevD.98.030001}{\doi{10.1103/PhysRevD.98.030001}}.

\bibitem{IceCube}
\hrefCMSnoop {}{{{I}ce{C}ube} Collaboration, ``Measurement of the multi-{T}ev
  neutrino interaction cross-section with {I}ce{C}ube using {E}arth
  absorption'',} \textit{ Nature} \textbf{ 551} (2017) 596,
  \href{http://dx.doi.org/10.1038/nature24459}{\doi{10.1038/nature24459}},
  \href{http://www.arXiv.org/abs/1711.08119}{\texttt{arXiv:1711.08119}}.

\bibitem{LEP}
\hrefCMSnoop {}{{ALEPH and DELPHI and L3 and OPAL and SLD Collaborations and
  LEP Electroweak Working Group and SLD Electroweak Group and SLD Heavy Flavour
  Group}, ``{Precision electroweak measurements on the Z resonance}'',}
  \textit{ Phys.Rept.} \textbf{ 427} (2006) 257,
  \href{http://dx.doi.org/10.1016/j.physrep.2005.12.006}{\doi{10.1016/j.physrep.2005.12.006}},
  \href{http://www.arXiv.org/abs/hep-ex/0509008}{\texttt{arXiv:hep-ex/0509008}}.

\bibitem{HFLAV}
\hrefCMSnoop {}{{Heavy Flavor Averaging Group, Y. Ahmis et al.}} \textit{
  Eur.Phys.J.} \textbf{ C77} (2017) 895,
  \href{http://dx.doi.org/10.1140/epjc/s10052-017-5058-4}{\doi{10.1140/epjc/s10052-017-5058-4}},
  \href{http://www.arXiv.org/abs/1612.07233v3}{\texttt{arXiv:1612.07233v3}}.

\bibitem{Pythia}
T.~Sj{\"o}strand\hrefCMSnoop {}{ {et~al.}, ``{An Introduction to PYTHIA
  8.2}'',} \textit{ Comput. Phys. Commun.} \textbf{ 191} (2015) 159,
  \href{http://dx.doi.org/10.1016/j.cpc.2015.01.024}{\doi{10.1016/j.cpc.2015.01.024}},
  \href{http://www.arXiv.org/abs/1410.3012}{\texttt{arXiv:1410.3012}}.

\bibitem{Fluka1}
\hrefCMSnoop {}{{T.T. Bohlen, F. Cerutti, M.P.W. Chin, A. Fass\`o, A. Ferrari,
  P.G. Ortega, A. Mairani, P.R. Sala, G. Smirnov, and V. Vlachoudis }, ``{The
  FLUKA Code: Developments and Challenges for High Energy and Medical
  Applications }'',} \textit{ Nuclear Data Sheets} \textbf{ 120} (2014)
  211--214.

\bibitem{Fluka2}
\hrefCMSnoop {}{G.~Battistoni {et~al.}, ``{Overview of the FLUKA code }'',}
  \textit{ Annals of Nuclear Energy} \textbf{ 82} (2015) 10--18.

\bibitem{DPMJET}
\hrefCMSnoop {}{A.~Fedynitch, ``{Cascade equations and hadronic interactions at
  very high energies}'',} \textit{ CERN-THESIS-2015-371
  {\url{https://cds.cern.ch/record/2231593}}} (27/11/2015).

\bibitem{LHCsim}
\hrefCMSnoop {}{A.~Lechner {et~al.}, ``{Validation of energy deposition
  simulations for proton and heavy ion losses in the CERN Large Hadron
  Collider}'',} \textit{ Physical Review Accelerators and Beams} \textbf{ 22}
  (2019) 071003.

\bibitem{Cerutti1}
\hrefCMSnoop {}{K.~R\/oed, M.~Brugger, and G.~Spiezia, ``An overview of the
  radiation environment at the {LHC} in light of {R2E} irradiation
  activities'',} \textit{ CERN-ATS-Note-2011-077 TECH
  {\url{https://cds.cern.ch/record/1382083}}} ({May 9, 2012}).

\bibitem{Cerutti2}
\hrefCMSnoop {}{R.~G. Al\'ia {et~al.}, ``{LHC} and {HL-LHC:} {P}resent and
  {F}uture {R}adiation {E}nvironment in the {H}igh-{L}uminosity {C}ollision
  {P}oints and {RHA} {I}mplications'',} \textit{ {IEEE TRANSACTIONS ON NUCLEAR
  SCIENCE}} \textbf{ 65} ({January 2018}) 448--456.

\bibitem{radmon1}
\hrefCMSnoop {}{G.~Spiezia {et~al.}, ``{A New RadMon Version for the LHC and
  its Injection Lines}'',} \textit{ IEEE Trans. Nucl. Sci.} \textbf{ 61} (2014)
  3424--3431.

\bibitem{radmon2}
\hrefCMSnoop {}{D.~Kramer {et~al.}, ``{LHC RadMon SRAM detectors used at
  different voltages to determine the thermal neutron to high energy hadron
  fluence ratio}'',} \textit{ IEEE Trans. Nucl. Sci.} \textbf{ 58} (2011)
  1117--1122.

\bibitem{CERN-EN}
\hrefCMSnoop {}{{Courtesy of CERN Engineering Department}}.

\bibitem{cosmics}
\hrefCMSnoop {}{E.~Bugaev {et~al.}, ``{A}tmospheric {M}uon {F}lux at {S}ea
  {L}evel, {U}nderground and {U}nderwater'',} \textit{ Phys.Rev.} \textbf{ D58}
  (1998) 05401,
  \href{http://www.arXiv.org/abs/hep-ph/9803488v3}{\texttt{arXiv:hep-ph/9803488v3}}.

\bibitem{donut}
\hrefCMSnoop {}{{{DONuT}} Collaboration, ``Final tau-neutrino results from the
  {DON}u{T} experiment'',} \textit{ Phys. Rev.} \textbf{ D78} (2008) 052002,
  \href{http://dx.doi.org/10.1103/PhysRevD.78.052002}{\doi{10.1103/PhysRevD.78.052002}},
  \href{http://www.arXiv.org/abs/0711.0728}{\texttt{arXiv:0711.0728}}.

\bibitem{DsTau}
\hrefCMSnoop {}{{The DsTau collaboration, Aoki, S., Ariga, A.,Ariga, T. et
  al.}, ``{DsTau: study of tau neutrino production with 400 GeV protons from
  the CERN-SPS}'',} \textit{ JHEP} \textbf{ 1} (2020) 33,
  \href{http://dx.doi.org/10.1007/JHEP01(2020)033}{\doi{10.1007/JHEP01(2020)033}}.

\bibitem{OPERA_MCSrec}
\hrefCMSnoop {}{{OPERA Collaboration, N. Agafonova et al.}, ``{Momentum
  measurement by the multiple Coulomb scattering method in the OPERA
  lead-emulsion target}'',} \textit{ New Journal of Physics} \textbf{ 14}
  (2012) 013026.

\bibitem{OPERA_EMrec}
\hrefCMSnoop {}{{F. Juget (OPERA Collaboration)}, ``{Electromagnetic shower
  reconstruction with emulsion films in the OPERA experiment}'',} \textit{ J.
  Phys: Conf. Ser.} \textbf{ 160} (2009) 012033,
  \href{http://dx.doi.org/10.1088/1742-6596/160/1/012033}{\doi{10.1088/1742-6596/160/1/012033}}.

\bibitem{SHIP_EMrec}
\hrefCMSnoop {}{S.~Shirobokov {et~al.}, ``{Machine Learning for electromagnetic
  showers reconstruction in emulsion cloud chambers}'',} \textit{ J. Phys:
  Conf. Ser.} \textbf{ 1085} (2018) 042025,
  \href{http://dx.doi.org/10.1088/1742-6596/1085/4/042025}{\doi{10.1088/1742-6596/1085/4/042025}}.

\bibitem{LHCb_charged}
\hrefCMSnoop {}{{{LHCb}} Collaboration, ``{M}easurement of charged particle
  multiplicities in $pp$ collisions at $\sqrt{s}$$=$ 7 {T}ev in the forward
  region'',} \textit{ Eur. Phys. J.} \textbf{ C 72} (2012) 1947,
  \href{http://dx.doi.org/10.1140/epjc/s10052-012-1947-8}{\doi{10.1140/epjc/s10052-012-1947-8}}.

\bibitem{TOTEM_charged}
\hrefCMSnoop {}{G.~Antchev {et~al.}, ``{Measurement of the forward charged
  particle pseudorapidity density in pp collisions at sqrt{s} = 7 TeV with the
  TOTEM experiment}'',} \textit{ EPL} \textbf{ 98} (2012) 31002,
  \href{http://www.arXiv.org/abs/1205.4105v1}{\texttt{arXiv:1205.4105v1}}.

\bibitem{derujula}
\hrefCMSnoop {}{A.~De~R\`ujula, E.~Fernandez, and J.~J. G\`omez-Cadenas,
  ``Neutrino fluxes at future hadron colliders'',} \textit{ Nucl. Phys.}
  \textbf{ B405} (1993) 80--108.

\bibitem{Park}
\hrefCMSnoop {}{H.~Park, ``{T}he estimation of neutrino fluxes produced by
  proton-proton collisions at $\sqrt{s}$=14 {T}e{V} of the {LHC}'',} \textit{
  J. High Energy Phys.} \textbf{ JHEP10} (2011) 092,
  \href{http://dx.doi.org/10.1007/JHEP10(2011)092}{\doi{10.1007/JHEP10(2011)092}},
  \href{http://www.arXiv.org/abs/arXiv:1110.1971v2}{\texttt{arXiv:arXiv:1110.1971v2}}.

\bibitem{FASERnu}
\hrefCMSnoop {}{{{FASER}} Collaboration, ``{Detecting and Studying High-Energy
  Collider Neutrinos with FASER at the LHC}'',} \textit{ Eur. Phys. J.}
  \textbf{ C80} (2020) 61,
  \href{http://dx.doi.org/10.1140/epjc/s10052-020-7631-5}{\doi{10.1140/epjc/s10052-020-7631-5}}.

\bibitem{nutauN}
\hrefCMSnoop {}{Y.~Jeong and M.~Reno, ``{T}au neutrino and antineutrino cross
  sections'',} \textit{ Phys. Rev.} \textbf{ D82} (2010) 033010.

\end{thebibliography}\endgroup

\end{document}